\documentclass[acmtog]{acmart}

\AtBeginDocument{%
  \providecommand\BibTeX{{%
    \normalfont B\kern-0.5em{\scshape i\kern-0.25em b}\kern-0.8em\TeX}}}

\setcopyright{acmlicensed}
\copyrightyear{2024}
\acmYear{2024} \acmVolume{43} \acmNumber{6} \acmArticle{219} \acmMonth{12}\acmDOI{10.1145/3687968}
\usepackage{hyperref}

\hypersetup{
    colorlinks=true,
    linkcolor=blue,
    urlcolor=blue
}

\acmConference[SIGGRAPH Asia '24]{SIGGRAPH Asia '24}{December 03--06 2024}{Tokyo, JPN}

\acmISBN{978-1-4503-XXXX-X/18/06}

\acmSubmissionID{papers\_852}

\usepackage{multirow}
\usepackage{graphicx}
\usepackage{booktabs}
\usepackage{wrapfig}
\usepackage{gensymb}
\usepackage{amsmath}
\usepackage{colortbl}
\usepackage{overpic}
\usepackage{xcolor}
\definecolor{myred}{RGB}{151,0,0}
\definecolor{myblue}{RGB}{0,0,140}
\definecolor{mygreen}{RGB}{0,101,0}

\newcommand{\myred}[1]{\textcolor{myred}{#1}}
\newcommand{\mygreen}[1]{\textcolor{mygreen}{#1}}
\newcommand{\myblue}[1]{\textcolor{myblue}{#1}}
\newcommand{\blue}[1]{\textcolor{blue}{#1}}

\newcommand*\diff{\mathop{}\!\mathrm{d}}

\begin{document}

\title{Deformation Recovery: Localized Learning for Detail-Preserving Deformations}
\author{Ramana Sundararaman}
\email{sundararaman@lix.polytechnique.fr}
\orcid{0000-0002-9303-399X}
\affiliation{%
  \institution{LIX, Ecole Polytechnique}
  \streetaddress{}
  \city{Paris}
  \state{}
  \country{France}
  \postcode{91120}
}
\author{Nicolas Donati}
\email{nicolas.donati@polytechnique.edu}
\orcid{0009-0005-8121-7483}
\affiliation{%
  \institution{Ansys}
  \streetaddress{}
  \city{Paris}
  \state{}
  \country{France}
}
\author{Simone Melzi}
\email{melzi@di.uniroma1.it}
\orcid{0000-0003-2790-9591}
\affiliation{%
  \institution{University of Milano-Bicocca}
  \streetaddress{}
  \city{Milan}
  \state{}
  \country{Italy}
}
\author{Etienne Corman}
\email{etienne.corman@inria.fr}
\orcid{0009-0002-9401-2362}
\affiliation{%
  \institution{Université de Lorraine, CNRS, Inria}
  \streetaddress{}
  \city{LORAINE}
  \state{}
  \country{France}
}
\author{Maks Ovsjanikov}
\email{maks@polytechnique.fr}
\orcid{0000-0002-5867-4046}

\affiliation{%
  \institution{LIX, Ecole Polytechnique}
  \streetaddress{}
  \city{Paris}
  \state{}
  \country{France}
  \postcode{91120}
}

\renewcommand{\shortauthors}{Sundararaman, et al.}

\begin{abstract}
We introduce a novel data-driven approach aimed at designing high-quality shape deformations based on a coarse localized input signal. Unlike previous data-driven methods that require a global shape encoding, we observe that detail-preserving deformations can be estimated reliably without any global context in certain scenarios. Building on this intuition, we leverage Jacobians defined in a one-ring neighborhood as a coarse representation of the deformation. Using this as the input to our neural network, we apply a series of MLPs combined with feature smoothing to learn the Jacobian corresponding to the detail-preserving deformation, from which the embedding is recovered by the standard Poisson solve. Crucially, by removing the dependence on a global encoding, every \textit{point} becomes a training example, making the supervision particularly lightweight. Moreover, when trained on a class of shapes, our approach demonstrates remarkable generalization across different object categories. Equipped with this novel network, we explore three main tasks: refining an approximate shape correspondence, unsupervised deformation and mapping, and shape editing. Our code is made available at \href{https://github.com/sentient07/LJN}{https://github.com/sentient07/LJN}.

\end{abstract}



\begin{teaserfigure}
\hspace{0.1cm}
  \begin{overpic}
  [width=0.99\textwidth] 
    {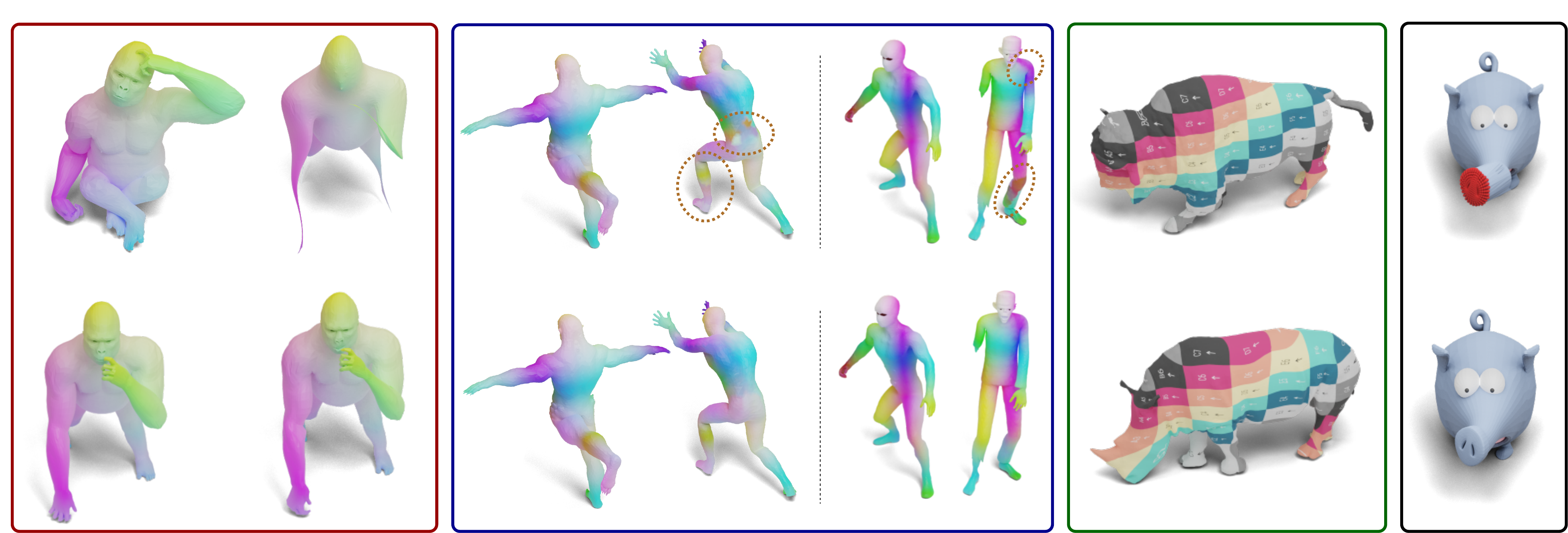}
    \put(-0.8,20.2){\rotatebox{90}{\textbf{\footnotesize Our input}}}
    \put(-0.8,4.2){\rotatebox{90}{\textbf{\footnotesize Our output}}}
    \put(14.6,21.4){\textbf{\LARGE $+$}
     }
    \put(4.7,16.5){\footnotesize Input Shape}
    \put(20.4, 17.9){\footnotesize Target}
    \put(16.7, 16.5){\footnotesize Spectral Projection}
    \put(5.7,0.8){\footnotesize Target}
    \put(22.7, 2.2){\footnotesize Our}
    \put(20.0, 0.8){\footnotesize Deformation}
     \put(10.1,33.05){\textbf{\footnotesize \myred{Deformation}}
     }
     \put(40.6,33.05){\textbf{\footnotesize \myblue{Refining Correspondence}}
     }
     \put(38.7,16.5){\footnotesize Input Correspondences with Errors}
     \put(40.7,0.8){\footnotesize Our Refined Correspondences}
     \put(71.0,33.05){\textbf{\footnotesize \mygreen{Unsupervised Matching}}}
     \put(71.0,16.5){\footnotesize Input Shape with Texture}
     \put(75.2,0.8){\footnotesize Our Matching}
     \put(92.1,33.05){\textbf{\footnotesize Editing}}
     \put(91.1, 17.9){\footnotesize Input Shape}
     \put(90.1, 16.5){\footnotesize $+$ Handles (red)}
    \put(91.8, 2.2){\footnotesize Our edited}
    \put(93.4, 0.8){\footnotesize Shape}
     \end{overpic}     
     \caption{\label{fig:teaser} We introduce a novel data-driven approach to produce high-quality shape deformations. Conditioned on a coarse representation of the deformation as the input signal (left top), we learn a detail-preserving deformation (right-bottom). 
     Leveraging this novel deformation framework, we perform three downstream tasks: refining an approximate shape correspondence, unsupervised shape matching, and interactive editing, as shown in subsequent columns.
     }
\end{teaserfigure}
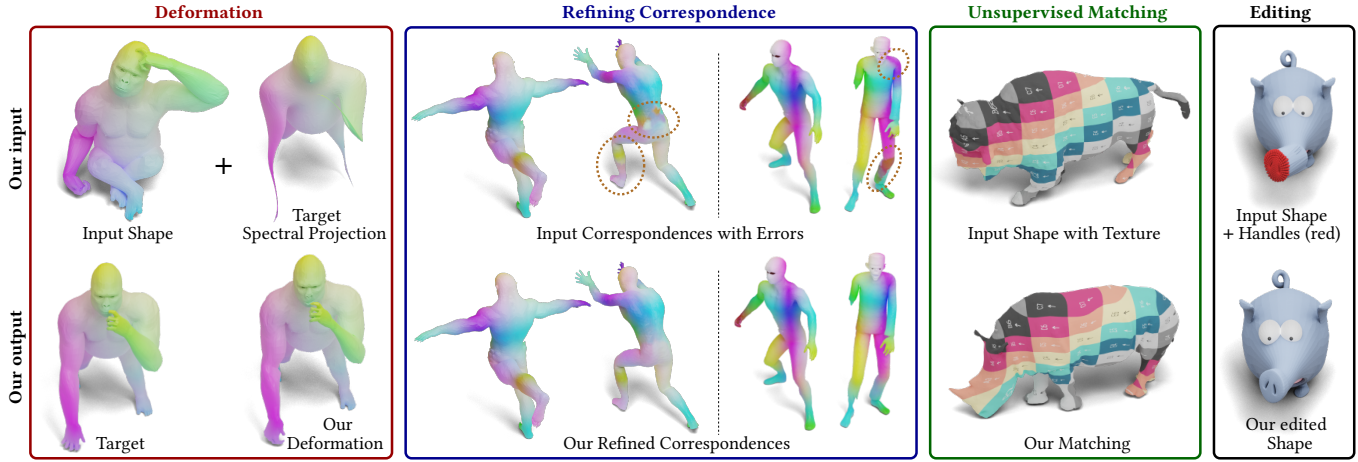


\maketitle

\section{Introduction}
Estimating meaningful deformations of surfaces is a classical problem in computer graphics~\cite{Terzopoulos1987,bechmann1994space}, with applications in several downstream tasks such as surface mapping and registration, character reposing, and handle-based editing to name a few~\cite{AigermanInj,innmann2016volumedeform,TransMatch,SparseICP,NICP,Li2008}. Due to its broad applicability, numerous techniques have been developed to address this task~\cite{BotschSurvey}.

While early approaches relied on numerical techniques~\cite{BotschSurvey}, recent methods, following the ubiquitous trend, leverage structured data-driven priors to estimate plausible deformations~\cite{groueix2018,Sundararaman2022DeformBasis,aigerman2022neural,Dodik2023}. Despite significant ongoing efforts, this problem remains challenging. A key difficulty with most existing data-driven methods lies in constructing a feasible \textit{latent space}~\cite{maesumi2023explore}. This typically requires amassing a collection comprising all plausible deformations~\cite{groueix20183d,aigerman2022neural}, thus necessitating a significant amount of training data. Furthermore, even when such data is available, representing shapes using a \textit{global encoding} typically ties existing approaches to specific \textit{shape categories}, and thus requires extensive re-training for cross-category generalization.

To that end, instead of relying on a global shape encoding, we condition our predictions on a \emph{coarse}, local deformation input signal. This choice is motivated by the observation that, in specific scenarios, as we will demonstrate in this paper, constructing an appropriate local input signal is sufficient to learn category-agnostic high-quality deformations. 
Building on the recent success of Jacobian fields as learnable deformation representation~\cite{aigerman2022neural}, we design a network that predicts Jacobian matrices per simplex. The input to our network is thus a field of Jacobian matrices, averaged to vertices around a one-ring neighborhood, representing a coarse deformation. Our network comprises a series of MLPs coupled with a smoothing operator, which takes coarse Jacobians as input and produces detailed Jacobians. Unlike the globally-conditioned approach in~\cite{aigerman2022neural}, our network has fully shared weights and is applied \textit{independently} at each simplex. 

Our network prediction is trained by supervising Jacobians corresponding to the detailed mesh, mesh vertex positions, and an integrability loss on Jacobians. Due to the purely local and fully shared nature of our networks, they can be trained using a handful of shapes, as each \textit{simplex} becomes a training instance. In summary, our deformation framework learns to \emph{produce} the input deformation at each vertex, represented as Jacobians, without utilizing any global information about the shape. For this reason, we refer to our network as the \emph{Local Jacobian Network} (LJN).

Given this general framework, the key question lies in the choice of the appropriate coarse input signal. In this work, we consider two general scenarios: shape correspondence and shape editing. For map refinement and shape correspondence, given a fixed source shape and a target shape representing a \emph{coarse} deformation of the source, we first express the coordinate function of the target shape using the low-frequency eigenfunctions of its Laplace-Beltrami Operator (LBO) to obtain a smooth approximation of its geometry~\cite{LevySpec}, referred to as the \emph{spectrally projected} shape. This is a natural representation supported by shape correspondence methods that use the functional map framework \cite{ovsjanikov2012functional}. 
We then define the Jacobians from the source shape to the spectrally projected target shape as the coarse deformation signal, which forms the input to our network. For shape editing, we use the rotation matrices at each face, averaged to the incident vertex, as the input signal to produce Jacobians corresponding to a valid shape. The network, training data, and supervision remain consistent across all input signals.

Since LJN relies on localized input signals, unlike a globally shape-aware network, it is inherently agnostic to shape categories~\cite{GuerreroEtAlPCPNetEG2018}. As a result, there is no need to amass a large training dataset or perform category-specific training, since, as mentioned above, each point (i.e., local region in the shape) becomes a training sample. Therefore, we limit our dataset to only 60 shape pairs in all our supervised training experiments. Trained on these 60 pairs of human shapes, LJN demonstrates remarkable generalization across object categories in recovering minimal distortion embeddings. 

We explore three main \textit{tasks} using our framework: map refinement, unsupervised non-isometric shape correspondence, and interactive shape editing as shown in Figure~\ref{fig:teaser}. In the first task, given an approximate vertex-based correspondence as input, we refine it using our deformation method. Specifically, we construct a coarse input signal by projecting the target geometry onto the spectral basis, from which we first recover the detail-preserving deformation using LJN. Subsequently, we perform an NNSearch in $\mathbb{R}^3$ to obtain an improved point-wise map (i.e., registration). Evaluated across standard near-isometry and non-isometry benchmarks, LJN achieves significant improvements in accuracy, coverage, map smoothness, and reduction of map inversions compared to existing map-refinement techniques, all while being non-iterative and fully differentiable. 

In our second task, we do not assume an input approximate correspondence but rather simultaneously estimate the map and the shape deformation. For this, we train LJN alongside a Deep Functional Map (DFM)~\cite{donati2022DeepCFMaps,sun2023spatially,ULRSSM} network on a collection of animal shapes with distinct mesh connectivities from the SMAL~\cite{li2022attentivefmaps} and SHREC-20~\cite{DykeShrec20} datasets. Our joint network improves map coverage and map smoothness compared to current state-of-the-art unsupervised correspondence techniques.

Finally, as our third task, to demonstrate that LJN can generalize to different local input signals, we focus on interactive shape editing. Specifically, a user deforms a shape by displacing selected vertices, and our goal is to solve for an embedding with minimal distortion. In this setup, we utilize the closest rotation matrix to the prescribed deformation as the input signal and learn to produce Jacobians corresponding to a valid shape. At the inference time, this task is akin to ARAP~\cite{Sorkine2006}, however, instead of iteratively updating the Jacobians, we learn a Jacobian that corresponds to detail-preserving deformation in a single feed-forward pass. We demonstrate that LJN generalizes to arbitrary object categories, producing more plausible and minimal distortion embeddings in comparison to ARAP.

\section{Related Works}
Shape deformation is a thoroughly researched yet continually evolving field. Given the extensive literature, we refer the readers to existing surveys~\cite{BotschSurvey}, and we focus our discussion on aspects directly relevant to our work.

\subsection{Neural Deformation Techniques}
In recent years, data-driven deformation techniques have shown promising results in shape registration and correspondence tasks~\cite{groueix20183d,Deprelle2019LearningES,TransMatch,Sundararaman2022DeformBasis}. Earlier models primarily learned displacement fields over a fixed template~\cite{Kanazawa2015Learning3D,groueix20183d,Litany2017DeformableSC}, relying on a global latent code derived from point-based~\cite{qi2016pointnet}, graph-based~\cite{Wang2019}, or mesh-based representations~\cite{hanocka2019meshcnn}, as well as implicit surfaces~\cite{bhatnagar2020loopreg}. The proliferation of extensive training datasets~\cite{varol17_surreal,STAR:2020} has facilitated the development of more sophisticated, template-free methods~\cite{TransMatch,aigerman2022neural}. Although neural deformation techniques predominantly focus on learning \textit{displacement fields}, several alternatives have emerged, including cage-based~\cite{Yifan:NeuralCage:2020,Dodik2023}, control-point-based~\cite{Kurenkov2017DeformNetFD,Sundararaman2022DeformBasis}, vector-field-based~\cite{jiang2020shapeflow}, and differential-based methods~\cite{aigerman2022neural}. Notably, the Neural Jacobian Field approach~\cite{aigerman2022neural} stands out due to efficiency and capacity to simulate realistic deformations. Consequently, this representation has been applied to various downstream applications, such as text and image-guided deformation~\cite{Gao_2023_SIGGRAPH, yoo2024apap}. However, to the best of our knowledge, nearly all data-driven deformation techniques utilize a \textit{global encoder} for learning across collections. While these methods are discretization-agnostic~\cite{groueix20183d,aigerman2022neural}, they often fail to generalize to unseen object categories or necessitate category-specific training. Additionally, they require a significant number of training shapes that share a common 1:1 correspondence. To overcome these limitations, our work employs a localized input signal for supervised and unsupervised training, thereby rendering our approach more data-efficient and category-agnostic.

\subsection{Map-Refinement}
Map-refinement techniques are typically iterative and can be broadly classified into two categories: spatial and spectral refinement techniques. The former considers the embedding of shapes in Euclidean space, while the latter focuses on the spectral domain, spanned by the first k-eigenfunctions of the Laplace-Beltrami Operator~\cite{Dubrovina2010MatchingSB}. A classical approach in the spatial domain includes Iterative Closest Point (ICP)~\cite{Chen1992,registration} and its specific adaptations~\cite{NICP,SparseICP,PotmannRegistration}. While the aforementioned approaches treat individual points separately,  probabilistic approaches such as Gaussian Mixture Models (GMMs)~\cite{CPD,BCPD} and Optimal Transport~\cite{Mandad2017Mapping,Solomon2016} have been well-explored. However, these methods struggle to converge when the shapes in the input pair are geometrically distant. 
Spectral approaches~\cite{ovsjanikov2012functional} are less affected by geometric proximity but often fail to recover high-frequency details due to the truncation of basis. To counteract this, upsampling techniques~\cite{MelziZO,Ren2018,Ren2021,li2022attentivefmaps} augment the spectral frequency during each iteration. Yet, their dependence on the alignment of intrinsic quantities limits generalization to significant non-isometries. While spatial coupling has been explored to address non-isometries~\cite{Ezuz2017,Ezuz2019,Ezuz2019b,magnet2022smooth}, these methods are costly and challenging to integrate into differentiable frameworks. In contrast, our approach achieves a more accurate point-wise map through a single feed-forward pass and a back-substitution being orders of magnitude faster.

\subsection{Functional Map and Deformation}
Corman et al.~\cite{Corman2017} were first to define distortion between pairs of meshes with differing connectivity using the functional map framework. Similarly to our approach, the authors proposed to pull back the intrinsic metric of the target shape onto the source. However, in that work, the authors used the shape difference operator~\cite{Rustamov2013} and constructed the embedding via Poisson solve~\cite{Panozzo:2014}. Since the Functional Map operator lacks extrinsic awareness, they employed an offset surface to fully recover the embedding. In contrast, our work involves pulling back the \textit{coordinate functions} in the truncated basis and \emph{learning} the Jacobian corresponding to the coordinate functions in the full basis. More recently, the basic deep functional map approach~\cite{litany2017deep,donati2020deepGeoMaps} has been enhanced with spatial awareness, often defined via properness~\cite{Ren2021} to enforce cycle consistency~\cite{ULRSSM,sun2023spatially}. Leveraging this observation, recent~\cite{jiang2023non} and concurrent work~\cite{Cao_2024_CVPR} have explored various deformation models to improve point-wise map extraction. Our work aligns with these developments, as our unsupervised deformation-mapping experiments demonstrated. Our deformation model is \emph{learned} alongside map estimation without requiring any correspondence information.

\section{Preliminaries}
\label{sec:Preliminaries}
%
\paragraph{Notations:}
We represent shapes as compact 2-dimensional Riemannian manifolds $\mathcal{M}$ possibly with boundary $\partial \mathcal{M}$. The space of square integrable functions on $\mathcal{M}$ is noted $\mathcal{L}^{2}\left( \mathcal{M} \right)$ and equiped with the scalar product $\left< f, g \right> = \int_\mathcal{M} fg \diff \mu$. We denote the tangent plane at $p$ as $T_{p} \mathcal{M}$. We discretize $\mathcal{M}$ as a triangle mesh $\mathcal{S} := \{\mathcal{V}, \mathcal{F}\}$ with $\mathcal{V}$  vertices and $\mathcal{F}$ faces. We refer to the list of the coordinates of the vertices $\mathcal{V} \in \mathbb{R}^{3}$ as the embedding of the shape.
On the manifold $\mathcal{M}$, we discretize the Laplace-Beltrami Operator (LBO), denoted $\Delta$, using the standard cotangent-based discretization~\cite{Pinkall1993ComputingDM}. We adopt the face-based discretization of the tangent plane~\cite{Azencot2013}. Since $\Delta$ is a symmetric operator, by solving the generalized eigenvalue problem $\Delta \Psi =\lambda M \Psi$, we obtain the LBO eigen-basis $\Psi=\left[\psi_1 \ldots \psi_k\right]$, where, $M$ is the diagonal-lumped vertex-mass matrix~\cite{Meyer2003}. These eigen-basis are an orthonormal basis (w.r.t. $M$) for the truncated subspace of $\mathcal{L}^{2}\left( \mathcal{M} \right)$  of smooth, low-frequency functions. Given an embedding $\mathcal{V}$, we consider its spectral projection$\bar{\mathcal{V}} = \Psi \Psi^{\dag}\mathcal{V}$, where $\Psi^{\dag}$ is the Moore-Penrose pseudo-inverse of $\Psi$. Analogously, we use $\bar{S}$ to denote the shape whose vertices are $\bar{\mathcal{V}}$. 
For each face, we define a possibly non-orthonormal frame $\mathcal{E} := [\vec{e}_1, \vec{e}_2, \vec{N}]^T$, which is a $3 \times 3$ matrix. The frame comprises the first two edge vectors of the face and the face normal. We denote the frames for all faces as $\mathbf{E} := [\mathcal{E}_1 \ldots \mathcal{E}_{|\mathcal{F}|}]$. 
For any given point $x \in \mathcal{M}$, a local deformation can be defined using a Jacobian matrix $J_{x} \in \mathbb{R}^{3 \times 3}$, the matrix of all the first-order partial derivatives of the deformation. In the discrete setting, given our face-based discretization of the tangent plane, Jacobians are piece-wise constant per-face. We denote $\mathbf{J} := [J_1 \ldots J_{|\mathcal{F}|}]$ to be the Jacobian across all faces represented in matrix form. 
If $x \in F_i$, we have $J_{x} = J_{i}$, where $F_i$ is the $i^{th}$ face. 

Given two discretized shapes $\mathcal{S}_1$ and $\mathcal{S}_2$ (often called source and target shapes) with respectively $n_1$ and $n_2$ vertices, we can write a correspondence $\varphi: \mathcal{S}_1 \rightarrow \mathcal{S}_2$ between those shapes as a binary matrix $\Pi \in \mathbb{R}^{n_1 \times n_2}$ where $\Pi[{i, j}] = 1$ denotes $j^{th}$ vertex on $\mathcal{S}_2$ being the image of $i^{th}$ vertex on $\mathcal{S}_1$. When the two shapes are in 1:1 correspondence, $\Pi_{12}$ is an identity matrix.

\paragraph{Functional Map}
The functional map pipeline, introduced in \cite{ovsjanikov2012functional}, is an efficient and compact representation for maps between shapes. More specifically, let $\varphi: \mathcal{S}_1 \rightarrow \mathcal{S}_2$ be a pointwise map, and $\Pi_{12}$ its corresponding binary matrix. The pull-back operator associated with this map, expressed in the LBO truncated eigen-basis and denoted as $\mathbf{C}_{21} \in \mathbb{R}^{k \times k}$, is referred to as the Functional Map~\cite{ovsjanikov2012functional}. $C_{21}: \mathcal{L}^{2}\left(S_2\right) \rightarrow \mathcal{L}^{2}\left(S_1\right)$ acts as a linear operator between the square integrable functions on the respective shapes. This can be derived from the binary matrix $\mathbf{C}_{21} = (\Psi_{1})^{\dag} \Pi_{12} \Psi_{2}$. Where, $\Psi_1^{\dag}$ is the Moore-Penrose pseudoinverse of $\Psi_1$. Given the functional map $C_{21}$, in the simplest setup, we can compute an associated $\Pi_{12}^*$ via a nearest-neighbors search~\cite{FMCourse}:

\begin{equation}
    \Pi_{12}^* = \underset{\Pi_{12}}{\operatorname{argmin}} \left\|\Psi_2 C_{21}- \Pi_{12} \Psi_1\right\|_F^2
\label{eqn:FM2P2P}
\end{equation}

\paragraph{Jacobian-based deformation}
Let $\mathcal{S}_1, \mathcal{S}_2$ be shapes with 1:1 corresponding vertices and same connectivity but two different vertex embeddings $\mathcal{V}_1, \mathcal{V}_2 \in \mathbb{R}^3$. We can consider the deformation between them as a per-vertex coordinate re-assignment. The linear part of this deformation quantifying the change in edge-vectors is referred to as the Jacobian $\mathbf{J}_{12}$ between $\mathcal{S}_1 \rightarrow \mathcal{S}_2$. This change of edge vectors ~\cite{Sumner,BotschPoisson}) is explicitly given by:  
\begin{equation}
    \mathbf{J}_{12} = \mathbf{E}_1^{-1} \mathbf{E}_2
\label{eqn:JacClosedForm}
\end{equation}
where $\mathbf{E}_\ell = [\mathcal{E}_1 \ldots \mathcal{E}_{|\mathcal{F}_\ell|}], \ \ell \in {1,2}$, are the frames corresponding to all faces, rewritten in matrix form. Note that while quantifying the change in edge vectors is sufficient to define $\mathbf{J}_{12}$, solving Eqn~\eqref{eqn:JacClosedForm} becomes under-determined (see Chap 3.1.2 \cite{Sumner2005MeshMU}). While~\cite{Sumner} overcomes this by tetrahedralization of each face, we simply use the \emph{unit-normal} vector, as changes along normal directions are inconsequential for a surface undergoing deformation in $\mathbb{R}^3$. Given a Jacobian $\mathbf{J}_{12}$, we can compute its closest  embedding in the least-square sense by solving the following Poisson equation:
\begin{equation}
    \Delta_1 \mathcal{V}_2 = \nabla_1^{T} A_1 \mathbf{J}_{12}
\label{eqn:Poisson}
\end{equation}

$\nabla_1^{T} A_1$ is the divergence operator where $\nabla_1$ is the gradient operator and $A_1$ is the area element of each face of $\mathcal{S}_1$. Since $\Delta_1$ is only semi-definite, the solution $V\mathcal{V}_2$ is not unique and is only valid up-to global translation \cite{BotschPoisson}. 

\section{Proposed Method}
 We begin by describing the general framework for training and performing inference with LJN in Section~\ref{subsec:General}. In Section~\ref{subsec:supervised}, we discuss the supervised training strategy for map refinement, followed by the inference strategy that alleviates the necessity for consistent connectivity in Section~\ref{subsec:Inference}. In Section~\ref{subsec:unsupervised}, we introduce a novel method to simultaneously learn the deformation and mapping between shape pairs with different triangulations. Finally, in Section~\ref{subsec:ShapeEditing}, we discuss the utility of LJN for handle-based shape editing.
\begin{figure}[t]
  \centering
  \includegraphics[width=\linewidth]{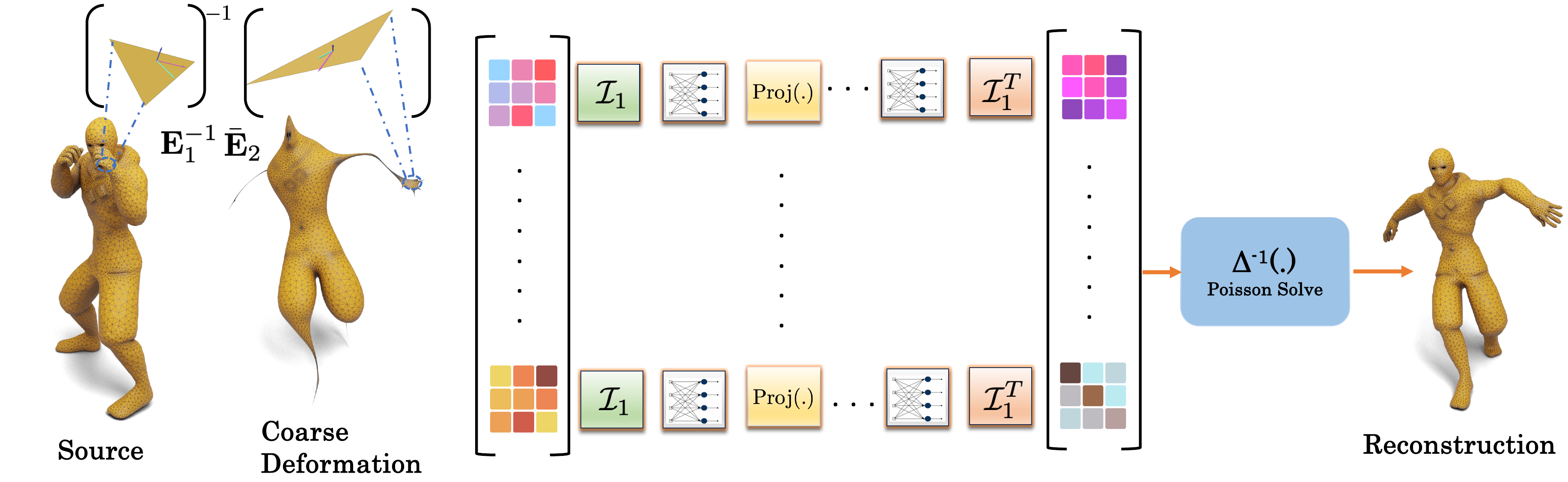}
  \caption{Illustration of our learning framework with spectrally projected input signal. Given a source shape and the coarse deformation signal, we feed in individual triangle deformations, as Jacobians, averaged to incident vertices. Then, we apply series MLPs coupled with spectral projection in the \emph{feature-space} to recover detail-preserving deformation.}
  \label{fig:Method}
\end{figure}

\subsection{Detailed Deformation Learning from Coarse Signals}
\label{subsec:General}
Given a source shape \(\mathcal{S}_1\) and a coarse deformation input signal, we aim to learn the detailed deformation of the source shape, expressed as Jacobian $\mathbf{J}_{12}$. Note that our coarse input signal might not correspond to a valid shape. When these Jacobians are integrated to produce an embedding, they must correspond to a plausible shape. We let \(\mathcal{S}_2\) be the shape which the target detailed Jacobian $\mathbf{J}_{12}$ corresponds to. We refer to \(\mathcal{S}_2\) as the target shape. We let $\Theta_{12} \in \mathbb{R}^{|\mathcal{V}|\times d }$ be the $d-$dimensional coarse input signal defined at vertex. In the case of Jacobians defined on faces, we first average them to vertices with the operator $\mathcal{I}$ mentioned below.
Recall that, unlike ~\cite{aigerman2022neural}, we do not rely on a global encoding of the target shape but instead condition the input to our network using $\Theta_{12}$ alone.

 \setlength{\columnsep}{4pt}
\begin{wrapfigure}[8]{r}{0.19\textwidth}
\vspace{-0.3cm}

\begin{overpic}[width=0.19\textwidth] 
    {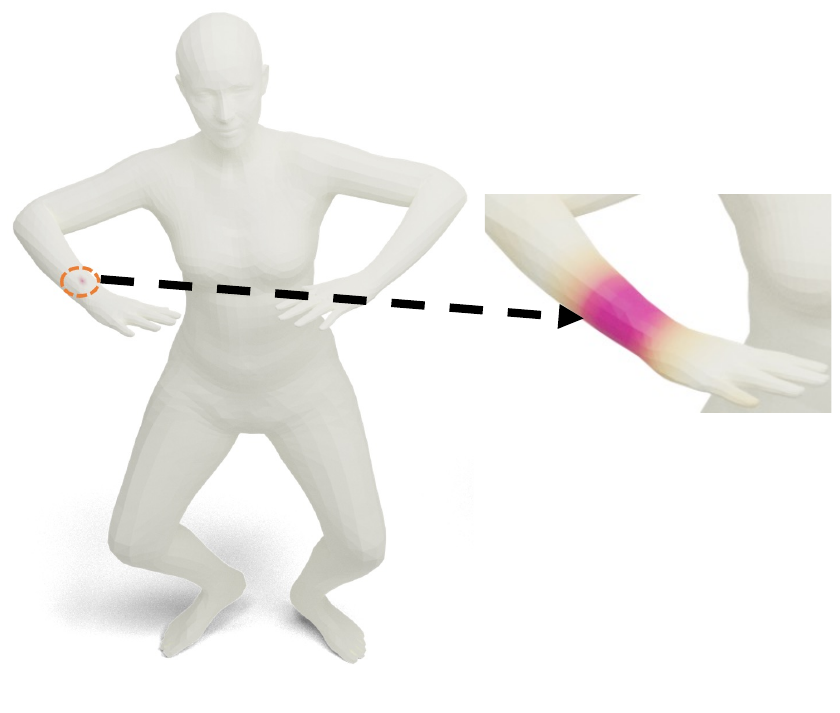}
    \put(56.6,78.4){\footnotesize Features after}
\put(66.6,71.4){\footnotesize spectral}
   \put(63.6,64.4){\footnotesize projection}     
     \end{overpic}
    \label{fig:Proj}
\end{wrapfigure}

 Given this input signal, we apply MLPs per-vertex in tandem with a \emph{spectral projection} layer to learn the detailed Jacobian corresponding to $\mathcal{S}_2$. The spectral projection layer projects the learned features to the eigenbasis of the LBO operator as given in Equation~\eqref{eqn:SpectralBlur}. We observed that this enables effective information sharing in a small local neighborhood around each vertex. For visualization purposes, we consider a single Jacobian as a Dirac signal defined at the $i^{th}$ vertex as shown in the inset figure (left). Then, upon projecting this signal to the eigenbasis of the LBO operator, the features are \emph{smoothly} spread out in a small neighborhood around the vertex of interest as shown in the inset figure (right). Our learning process can then be described as follows:
\begin{equation}
    \hat{\mathbf{J}}_{12} = \mathcal{I}_{1}^{T} \mathcal{G}_{6}( \ldots (\mathrm{Proj}_{1}( \mathcal{G}_{1}(\Theta_{12} ))) 
\end{equation}

The above equation sequentially applies an MLP, $\mathcal{G} (.)$, whose specifications are provided in the Supplementary. Here, $\mathrm{Proj}(\cdot) := \Psi_1 \Psi_1^T M_1 \cdot$ expresses features in the orthonormal basis defined by $\Psi_1$. The operator $\mathcal{I}_{1}$ is a sparse row-stochastic matrix that averages quantities from faces to vertices. The non-zero entries are corresponding to row $i$ of $\mathcal{I}_{1}$ sum up to 1 and are all equal to $\frac{1}{f_{i}}$, where $f_{i}$ denotes the number of faces incident on vertex $i$. Its transpose, $\mathcal{I}^{T}_{1}$ averages the per-vertex predictions back to faces. Given the ground-truth Jacobian $\mathbf{J}_{12}^{*}: \mathcal{S}_1 \rightarrow \mathcal{S}_2$, to train our network, we minimize a loss function consisting of three terms as follows:
\begin{equation}
    \mathcal{L}_{\mathrm{Tr}} = \alpha_1 \left|\left| \mathbf{J}_{12}^{*}
 - \hat{\mathbf{J}}_{12}\right|\right|_F^{2} + \alpha_2 \left|\left| \mathcal{V}_2 - \hat{\mathcal{V}}_1 \right|\right|_2^{2} + \alpha_3 \left|\left| \mathbf{J}_{12}^{*}
 - \mathbf{E}_1^{-1} \mathbf{\hat{E}} \right|\right|_F^{2}.
\end{equation}

Here $\hat{\mathcal{V}}_1 = \Delta_1^{-1} \nabla_1^{T} A_1 \hat{\mathbf{J}}_{12}$ is the embedding corresponding to the predicted Jacobian recovered via the Poisson solve (cf. Equation~\ref{eqn:Poisson}). The first two terms are used to supervise the positions and predicted Jacobians w.r.t the ground truth, while the third term penalizes discrepancies between the \emph{integrated} Jacobian and the ground truth Jacobian. More precisely, we compute the Jacobian corresponding to the recovered embedding $\hat{\mathcal{V}}_1$ using the corresponding non-orthonormal frame as $\mathbf{E}_1^{-1} \mathbf{\hat{E}}_1$ and minimize the discrepancy with the ground truth Jacobian. 

Our learning pipeline, along with one possible coarse input signal used in this paper, is visualized in Figure~\ref{fig:Method}. In summary, given some input Jacobians, we first average them to vertices, spectrally project them and then train our to reconstruct the detailed Jacobian between the source and target shapes. At inference time, the embedding can be obtained via a simple feedforward pass and a Poisson solve (cf. Eqn~\ref{eqn:Poisson}). Task-specific inference is discussed in subsequent sections.

\subsection{Supervised Learning from Spectral Inputs}
\label{subsec:supervised}
In this section, we discuss the construction of a coarse deformation input signal geared for shape registration-related tasks. Our choice of representation for the input signal is designed to be easily integrated into existing shape correspondence frameworks~\cite{ovsjanikov2012functional}.
To construct the input signal, we first express the coordinate function of the target shape using the first \(k\) eigenfunctions of the LBO, referred to as the \emph{spectrally projected} shape. Let $\bar{\mathcal{S}}_2$ be the shape obtained by applying spectral projection to $\mathcal{S}_2$, and $\bar{\mathcal{V}}_2$ be its vertex set. Then, the closed-form expression for $\bar{\mathcal{V}}_2$ is given as follows:
\begin{equation}
    \mathrm{Proj}(\mathcal{V}_2) = \bar{\mathcal{V}}_2 = \Psi_2 \Psi_2^{\dag} \mathcal{V}_2 = \Psi_2 \Psi_2^T M_2 \mathcal{V}_2
\label{eqn:SpectralBlur}
\end{equation}

Where $\Psi_2^\dag = \Psi_2^T M_2$ because of orthonormality with respect to $M_2$. Denoting the frames per face corresponding to the spectrally projected shape as $\bar{\mathbf{E}}_2$, the input signal to our network \textit{during training} is defined as follows:
\begin{equation}
\Theta_{12} := \mathcal{I}_{1} \mathbf{J}_{12} = \mathcal{I}_{1} \mathbf{E}_1^{-1} \bar{\mathbf{E}}_2
\label{eqn:JacInput}
\end{equation}

\begin{figure}[t]
  \centering
  \includegraphics[width=\linewidth]{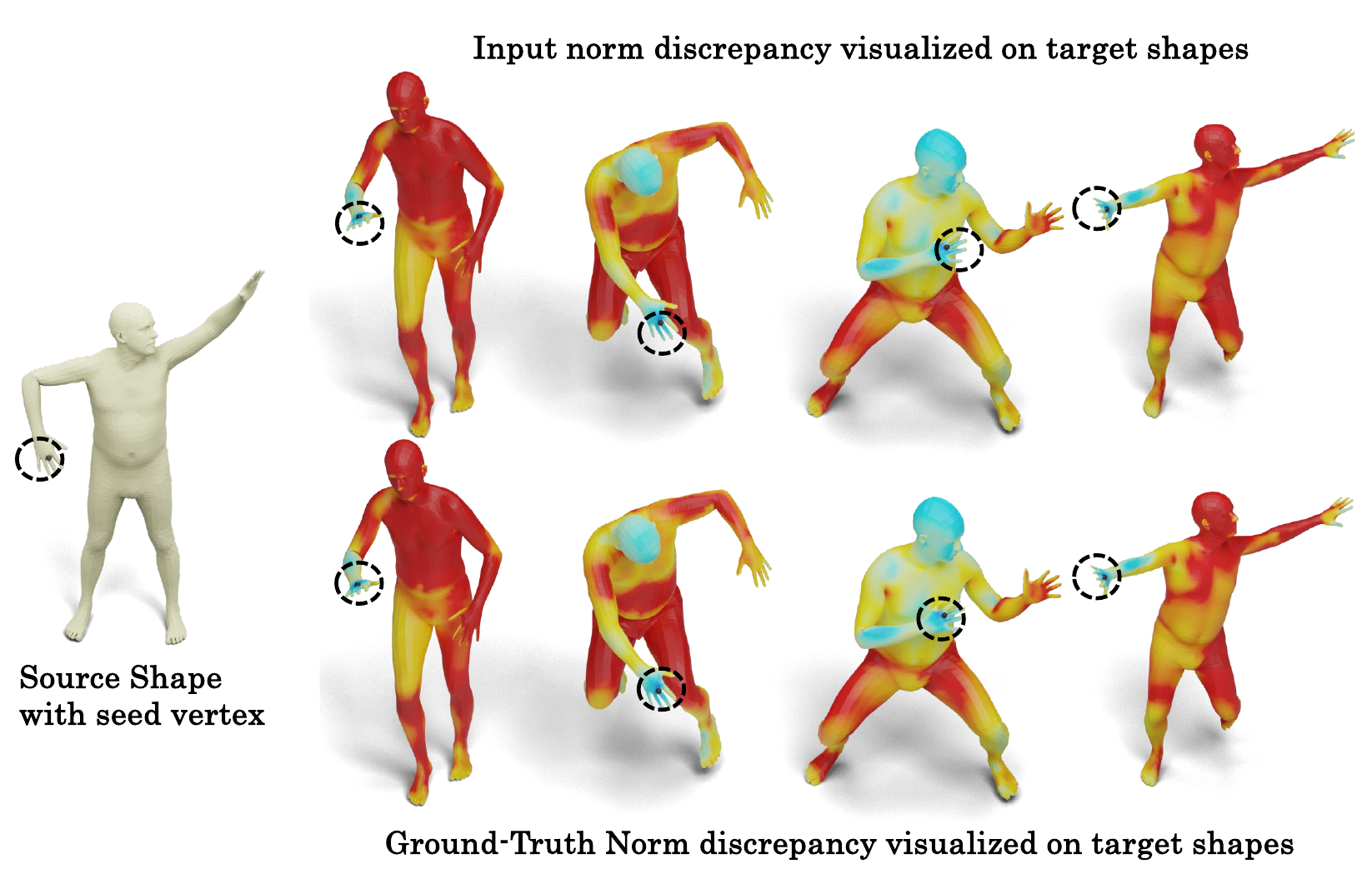}
  \caption{Illustration of the fidelity of our training framework. We select a \emph{seed} vertex, marked with a dotted circle. The Frobenius norm of the difference between the Jacobian at this seed vertex and all other vertices is plotted. The similarity in the distribution between the input (projected Jacobian) and the ground truth suggests the well-posed nature of our learning problem.}
  \label{fig:ApproachValidity}
\end{figure}

\begin{figure}[t]
  \centering
  \includegraphics[width=\linewidth]{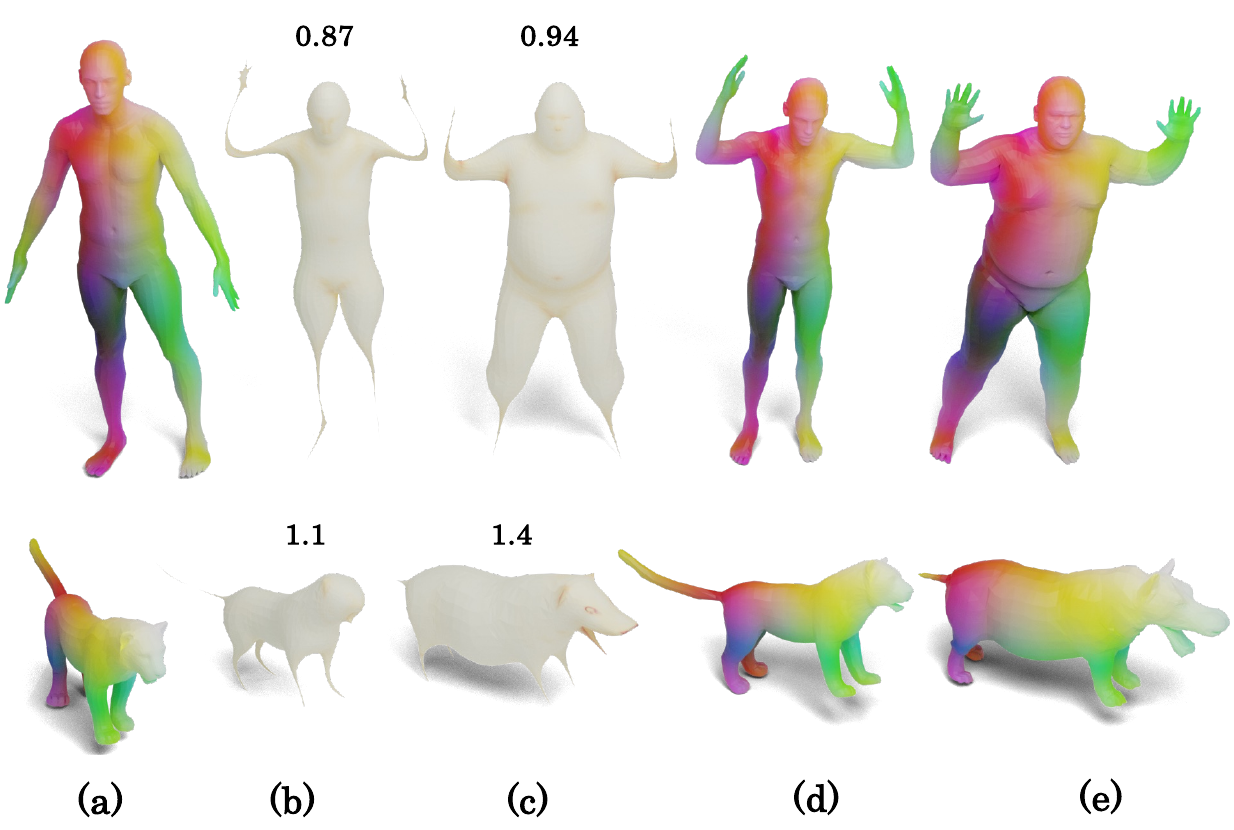}
  \vspace{-0.3cm}
  \caption{Visual insights on how our input signal is affected by near-isometric and non-isometric deformations. In the first row, we show a human shape undergoing two deformations into a similar pose but taken by two subjects, one similar and one not. We compare the discrepancy between the input Jacobian and the ground truth Jacobian as the Frobenius norm (plotted in the second and third columns). We repeat this for a pair of animals with a significantly higher degree of non-isometry between them. We observe that our input signal $\Theta_{12}$, remains comparable to the ground-truth across different levels of non-isometry.}
  \label{fig:IntvsExt}
\end{figure}

\paragraph{Discussion}
For a learning problem to be well-posed, distinct inputs should produce distinct outputs. We demonstrate this qualitatively on pairs of shapes, as shown in Figure~\ref{fig:ApproachValidity}. We consider a fixed source shape and a vertex of interest in this source shape as the "seed" vertex, say the $i^{th}$ vertex. We then plot the Frobenius norm of the difference between the input signal at the $i^{th}$ vertex and all the remaining vertices, expressed as $\|\Theta_{i, 12} - \Theta_{j, 12}\|_F \forall i \neq j$, in the first row. Similarly, in the second row, we plot the difference between the ground truth Jacobian $i^{th}$ vertex and all the remaining vertices, $\|J^{*}_{i, 12} - J^{*}_{j, 12}\|_F \forall i \neq j$. The similarity in the distributions of the norm at each vertex between input and ground truth across different deformations asserts the well-posed nature of our learning problem. Additionally, in Figure~\ref{fig:IntvsExt}, we provide an intuition on how this input signal behaves over near and non-isometric deformations.

\subsection{Inference: Meshes with differing connectivity.}
\label{subsec:Inference}
The requirement for shapes to be in 1-1 correspondence when constructing the input signal is impractical for real-world applications. To address this, we utilize the Functional Map framework to model deformations between shapes with differing connectivities. Therefore, given a pair of shapes $\mathcal{S}_1, \mathcal{S}_2$ with differing connectivity and a functional map $C_{21}$ between them, we aim to extract the deformation $\hat{\mathbf{J}}_{12}$ between them. The Functional Map can be estimated via an existing Deep Functional Map (DFM) framework~\cite{sun2023spatially,ULRSSM}, or using our unsupervised method detailed in Section~\ref{subsec:unsupervised}.

Different from training time, at inference, we first write the spectral projection of the coordinate function of the target shape and `pull' back the embedding function to the source shape using the functional map, as follows,

\begin{equation}
    \bar{\mathcal{V}}_1 = \Psi_1 C_{21} \Psi_2^{\dag} \mathcal{V}_2
\label{eqn:BlurInference}
\end{equation}

\begin{figure}[t]
  \centering
  \includegraphics[width=\linewidth]{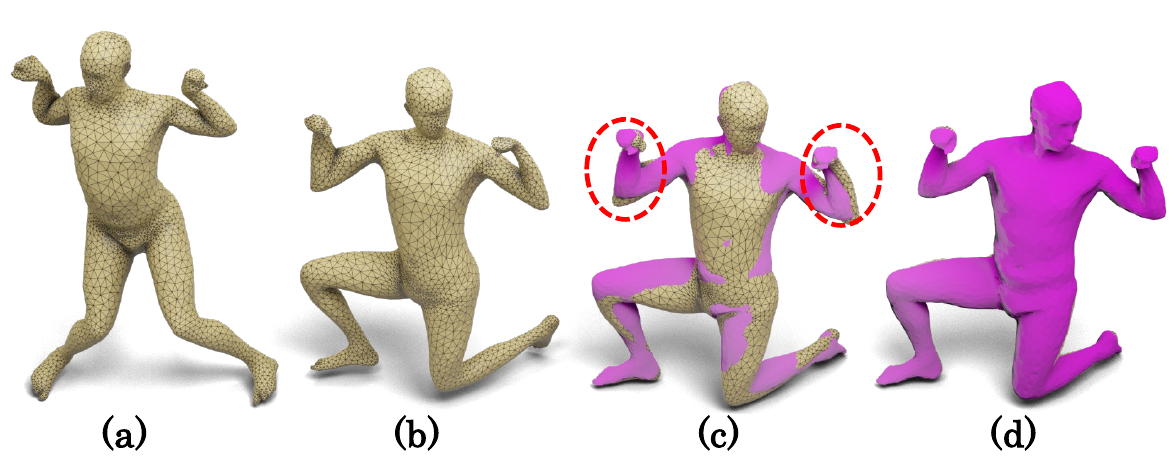}
  \vspace{-0.75cm}
  \caption{Given a source (a) and a deformation (Jacobian), we compute the embedding following Eqn~\ref{eqn:Poisson} and visualize in (b). While the deformation is near-perfect, the surfaces do not `\emph{align}' as shown in (c) where (b) is juxtaposed to the target shape. Computing the embedding via Eqn~\ref{eqn:EmbedRecoveryBacksub} yields a surface that is geometrically closer to the target as shown in (d).}
  \label{fig:UnAligning}
\end{figure}

The above equation expresses the coordinate function of the target shape in the LBO eigenbasis of the source shape, with the functional map acting as the change of basis operator. Since $\mathcal{V}_1$ and $\bar{\mathcal{V}}_1$ share the same connectivity, we can define the input to our network as given in Equation~\eqref{eqn:JacInput} and perform a feedforward pass to obtain $\hat{\mathbf{J}}_{12}$. 
To retrieve the embedding, instead of simply solving the Poisson equation~\eqref{eqn:Poisson}, we also leverage the point-wise map $\Pi_{12}$ arising from the Functional Map (using c.f Eqn~\eqref{eqn:FM2P2P}). We do this to address numerical errors in the learned Jacobian which might lead to detailed deformation but ``un-aligning'' surface as highlighted in Figure~\ref{fig:UnAligning}. To this end, our embedding recovery from Jacobian is solved via the following minimization:
\begin{equation}
    \hat{\mathcal{V}}_1 = \underset{\hat{\mathcal{V}}_1} {\operatorname{min}} \hspace{1mm} \alpha_4 \left|\left| \hat{\mathcal{V}}_1 - \Pi_{12} \mathcal{V}_2 \right|\right|^2_{M_{1}} + \alpha_5 \left|\left| \Delta_1 \hat{\mathcal{V}}_1 - \nabla_1^{T} A_{1} \hat{\mathbf{J}}_{12} \right|\right|^2_{M_{1}} + \left|\left| \hat{\mathcal{V}}_1 \right|\right|^2_{\Delta_{1}}
\label{eqn:EmbedRecovery}
\end{equation}

Here, the first term ensures that the recovered embedding remains geometrically close to the target surface, while the second term promotes the recovered embedding to respect the learned Jacobians. The final term encourages smoothness in the recovered embedding by favoring a minimum norm solution. Setting the gradient of the above expression w.r.t $\mathcal{V}_1$ to be zero yields:
\begin{equation}
    (\Delta_1 + \alpha_4 M_1 + \alpha_5 \Delta_1^T M_1 \Delta_1) \hat{\mathcal{V}}_1 = \alpha_4 M_1 \Pi_{12} \mathcal{V}_2 + \alpha_5 \Delta_1^T M_1 \nabla_1^{T} A_1 \hat{\mathbf{J}}_{12}
\label{eqn:EmbedRecoveryBacksub}
\end{equation}

Since the LHS in the above expression depends only on the Laplacian and the mass matrix, it can be pre-factored. If registration is the final goal, one can obtain a refined pointwise map using the deformed embedding $\hat{\mathcal{V}}_1$. This can be done by the following minimization:
\begin{equation}
    \Pi_{12}^{*} = \underset{\Pi^*_{12}}{\operatorname{argmin}} \| \hat{\mathcal{V}}_1 - \Pi^*_{12} \mathcal{V}_2 \|^2_2
\label{eqn:FinalP2P}
\end{equation}

which yields, $\Pi_{12}^{*} = \mathrm{NNSearch}(\hat{\mathcal{V}}_1, \mathcal{V}_2)$.

\subsection{Unsupervised Deformation Learning and Mapping}
\label{subsec:unsupervised}

In this section, we propose an unsupervised training methodology on a collection of shapes without consistent triangulation, leveraging the Functional Map framework. Unlike in Section~\ref{subsec:Inference}, we do not assume to have access $C_{21}$ to as input, but rather \emph{learn} it alongside $\mathbf{J}_{12}$. To this end, we employ a recent two-branch Deep Functional Map (DFM)~\cite{sun2023spatially,ULRSSM} approach and combine it with LJN. We denote the feature extractor within the standard DFM~\cite{DFN} corresponding to source and target shapes as $\mathcal{D}_1$ and $\mathcal{D}_2$ respectively. Between the input pair, we estimate the Functional Map in a differentiable manner, similar to ~\cite{ULRSSM} as follows,
\begin{equation}
    \hat{C}_{21} = \Psi_{1}^{\dag} \tilde{\Pi}_{12} \Psi_{2}
\label{eqn:SoftFmap}
\end{equation}
Where $\tilde{\Pi}_{12} = \operatorname{Softmax}\left(\mathcal{D}_{1} \mathcal{D}_{2}^T / \tau\right)$, is the soft pointwise map with $\tau=0.07$ being the temperature parameter. Using the $\hat{C}_{21}$ estimated above, we compute $\bar{\mathcal{V}}_1$ using Eqn~\ref{eqn:BlurInference} and use it to compute $\Theta_{12}$ via Eqn~\eqref{eqn:JacInput}. Then, we perform a feed-forward pass through LJN to get predicted Jacobian $\hat{\mathbf{J}}_{12}$. We train the two-branch functional map network and our LJN jointly by minimizing the following loss:
\begin{equation}
    \mathcal{L}_{\mathrm{un}} = \sum_{i\in\{1,2\}, j\in\{2,1\}} \left\|\mathbf{C}_{i,j}^T \mathbf{C}_{i,j}-\mathbf{I}\right\|_F^2+\left\|\mathbf{C}_{i,j}-\hat{\mathbf{C}}_{i,j}\right\|_F^2 + \mathcal{L}_{\mathrm{J}} (\mathbf{J}_{j,i})
\label{eqn:UnsupFMLoss}
\end{equation}

Where, $C_{i,j}$ is the predicted Functional Map and $\hat{C}_{i,j}$ is the functional map arising from the soft-pointwise map (c.f Eqn~\eqref{eqn:SoftFmap}). Note that the summation in the above equation optimizes the Functional Map and Jacobian in both directions. For simplicity and consistency, we elaborate on the minimization objective for $\mathbf{J}_{1,2}$ while noting that the same applies analogously to $\mathbf{J}_{2,1}$. Our unsupervised deformation objective $\mathcal{L}_{\mathrm{J}} (\mathbf{J}_{1,2})$ is given as follows:
\begin{equation}
    \mathcal{L}_{\mathrm{J}} (\hat{\mathbf{J}}_{12}) = \left\| \mathring{\mathbf{J}}_{12} - \hat{\mathbf{J}}_{12}\right\|^2_F + \alpha_6 \left\|\hat{\mathbf{J}}_{12} - \mathcal{H}_{1} \hat{\mathbf{J}}_{12}\right\|_F^2 + \alpha_7 \left\| \mathrm{det}(\hat{\mathbf{J}}_{12}) - 1. \right\|^2
\label{eqn:UnsupLoss}
\end{equation}

Here,  $\mathring{\mathbf{J}}_{12}$ is the Jacobian corresponding to the deformation prescribed by $\tilde{\Pi}_{12}$. More specifically, we first compute the deformation of the source embedding as $\mathring{\mathcal{V}_1} = \tilde{\Pi}_{12}\mathcal{V}_2$. Then, denoting the frame corresponding to mesh $\mathring{\mathcal{S}}_1 = \{\mathring{\mathcal{V}_1}, \mathcal{F}_1\}$ as $\mathring{\mathbf{E}}_1$, we define $\mathring{\mathbf{J}}_{12} := \mathbf{E}^{-1}_1 \mathring{\mathbf{E}}_1$. The second term in the loss function is a smoothness prior, enforcing minimal norm differences between predicted Jacobians corresponding to adjacent faces with $\mathcal{H}_{1}$ denoting the unsigned face-face incidence matrix. The third term promotes volume preservation, an important regularization term, which is deemed empirically useful when supervising with $\tilde{\Pi}_{12}$ due to the effect of truncation of basis.

\subsection{Interactive Editing}
\label{subsec:ShapeEditing}
To demonstrate that the LJN deformation framework is applicable to tasks beyond shape correspondence, we consider the task of interactive shape editing. In this scenario, a user prescribes deformation at selected vertices (handles). Our goal is to produce a minimal distortion embedding consistent with the user-prescribed deformation. Interactive editing can often result in meshes with degenerate elements, making the computation of differential operators and their eigendecomposition ill-defined. Therefore, we use the closest rotation matrix to the ground truth Jacobians as our input signal. This scenario is more challenging since the input deformation signal does not correspond to a valid shape, i.e, there is no guarantee that an embedding exists whose differentials correspond to the input signal.  
To compute the input signal, we apply polar decomposition to the ground truth Jacobian $\mathbf{J}^{*}_{12} = \mathcal{Q}_{12} \mathcal{W}_{12}$, where $\mathcal{Q}_{12}$ is the orthonormal matrix and $\mathcal{W}_{12}$ is an SPD matrix. Since our training pairs consist of valid shapes, $\mathbf{J}^{*}_{12}$ is always full rank and admits a unique polar decomposition. Therefore, to compute $\Theta_{12}$ (cf. Eqn~\ref{eqn:JacInput}), we use $\mathcal{Q}_{12}$ instead of $\mathbf{E}_1^{-1}\bar{\mathbf{E}}_2$. The rest of the training pipeline remains identical to our discussions in Section~\ref{subsec:supervised}. In summary, given rotation matrices averaged over the one-ring neighborhood, we learn the Jacobian corresponding to the detail-preserving deformation.

\begin{table*}[h]
\centering
\resizebox{\textwidth}{!}{%
\begin{tabular}{@{}|l|llllllllllll|llllllll|@{}}
\toprule
Refinement   & \multicolumn{12}{l|}{Near-Isometry}                                                                                                                                                                                                                                                                                                                                                                                         & \multicolumn{8}{l|}{Non-Isometry}                                                                                                                                                                                                                                           \\ \midrule
 & \multicolumn{4}{l|}{FAUST~\cite{Ren2018}}                                                                                                                     & \multicolumn{4}{l|}{SCAPE~\cite{Ren2018}}                                                                                                                     & \multicolumn{4}{l|}{SHREC-19~\cite{SHREC19}}                                                                                             & \multicolumn{4}{l|}{SMAL~\cite{li2022attentivefmaps}}                                                                                                                       & \multicolumn{4}{l|}{DT4D-Inter~\cite{magnet2022smooth}}                                                                                           \\ \midrule
  & \multicolumn{1}{l|}{Geod}         & \multicolumn{1}{l|}{Inv}          & \multicolumn{1}{l|}{DirE}         & \multicolumn{1}{l|}{Cov}           & \multicolumn{1}{l|}{Geod}         & \multicolumn{1}{l|}{Inv}          & \multicolumn{1}{l|}{DirE}         & \multicolumn{1}{l|}{Cov}           & \multicolumn{1}{l|}{Geod}         & \multicolumn{1}{l|}{Inv}          & \multicolumn{1}{l|}{DirE}         & Cov           & \multicolumn{1}{l|}{Geod}         & \multicolumn{1}{l|}{Inv}           & \multicolumn{1}{l|}{DirE}         & \multicolumn{1}{l|}{Cov}           & \multicolumn{1}{l|}{Geod}         & \multicolumn{1}{l|}{Inv}          & \multicolumn{1}{l|}{DirE}         & Cov           \\ \midrule
 Init   
    & \multicolumn{1}{l|}{\blue{1.7}}          & \multicolumn{1}{l|}{7.4}          & \multicolumn{1}{l|}{3.2}          & \multicolumn{1}{l|}{80.8}          & \multicolumn{1}{l|}{2.3}          & \multicolumn{1}{l|}{8.1}         & \multicolumn{1}{l|}{4.2}          & \multicolumn{1}{l|}{76.2}          & \multicolumn{1}{l|}{\blue{5.1}}          & \multicolumn{1}{l|}{\blue{7.9}}         & \multicolumn{1}{l|}{6.6}          & 72.6         & \multicolumn{1}{l|}{\blue{4.7}}          & \multicolumn{1}{l|}{13.6}          & \multicolumn{1}{l|}{13.9}         & \multicolumn{1}{l|}{62.0}          & \multicolumn{1}{l|}{6.8}          & \multicolumn{1}{l|}{9.8}         & \multicolumn{1}{l|}{17.8}           & 61.9          \\ \cmidrule(l){1-21} 
Spec Proj  
& \multicolumn{1}{l|}{3.1}          & \multicolumn{1}{l|}{11.4}         & \multicolumn{1}{l|}{3.1}          & \multicolumn{1}{l|}{58.5}          & \multicolumn{1}{l|}{3.0}          & \multicolumn{1}{l|}{9.9}          & \multicolumn{1}{l|}{3.2}          & \multicolumn{1}{l|}{58.3}          & \multicolumn{1}{l|}{5.8}          & \multicolumn{1}{l|}{10.5}         & \multicolumn{1}{l|}{3.5}          & 53.9          & \multicolumn{1}{l|}{4.9}          & \multicolumn{1}{l|}{\textbf{10.1}}          & \multicolumn{1}{l|}{\textbf{3.5}}          & \multicolumn{1}{l|}{58.0}          & \multicolumn{1}{l|}{7.4}          & \multicolumn{1}{l|}{10.5}         & \multicolumn{1}{l|}{4.4}          & 47.0          \\ \cmidrule(l){1-21} 
$\alpha_5$=0
& \multicolumn{1}{l|}{\blue{1.7}}          & \multicolumn{1}{l|}{\blue{7.2}}          & \multicolumn{1}{l|}{3.0}          & \multicolumn{1}{l|}{81.2}          & \multicolumn{1}{l|}{\blue{2.1}}          & \multicolumn{1}{l|}{\blue{5.7}}         & \multicolumn{1}{l|}{3.4}          & \multicolumn{1}{l|}{78.2}          & \multicolumn{1}{l|}{\blue{5.1}}          & \multicolumn{1}{l|}{\blue{7.9}}         & \multicolumn{1}{l|}{4.5}          & 72.9          & \multicolumn{1}{l|}{\blue{4.7}}          & \multicolumn{1}{l|}{14.2}          & \multicolumn{1}{l|}{7.5}          & \multicolumn{1}{l|}{62.9}          & \multicolumn{1}{l|}{6.6}          & \multicolumn{1}{l|}{\blue{9.4}}         & \multicolumn{1}{l|}{8.1}         & 63.1          \\ \cmidrule(l){1-21} 
ZO 
& \multicolumn{1}{l|}{\blue{1.7}}          & \multicolumn{1}{l|}{10.9}         & \multicolumn{1}{l|}{3.1}          & \multicolumn{1}{l|}{\blue{83.3}} & \multicolumn{1}{l|}{2.2}          & \multicolumn{1}{l|}{8.2}          & \multicolumn{1}{l|}{3.6}          & \multicolumn{1}{l|}{\blue{80.8}} & \multicolumn{1}{l|}{\blue{5.1}}          & \multicolumn{1}{l|}{9.7}          & \multicolumn{1}{l|}{4.0}          & \blue{74.2}          & \multicolumn{1}{l|}{5.6}          & \multicolumn{1}{l|}{11.9}          & \multicolumn{1}{l|}{24.2}         & \multicolumn{1}{l|}{\blue{69.9}}          & \multicolumn{1}{l|}{6.6}          & \multicolumn{1}{l|}{11.1}         & \multicolumn{1}{l|}{11.6}         & 69.7          \\ \cmidrule(l){1-21} 
DZO 
& \multicolumn{1}{l|}{1.8}          & \multicolumn{1}{l|}{10.4}         & \multicolumn{1}{l|}{3.2}          & \multicolumn{1}{l|}{79.8}          & \multicolumn{1}{l|}{2.4}          & \multicolumn{1}{l|}{10.4}         & \multicolumn{1}{l|}{3.9}          & \multicolumn{1}{l|}{78.3}          & \multicolumn{1}{l|}{\blue{5.1}}          & \multicolumn{1}{l|}{10.7}         & \multicolumn{1}{l|}{3.9}          & 71.6          & \multicolumn{1}{l|}{4.9}          & \multicolumn{1}{l|}{11.1}          & \multicolumn{1}{l|}{9.6}         & \multicolumn{1}{l|}{68.5}          & \multicolumn{1}{l|}{\blue{6.3}} & \multicolumn{1}{l|}{9.9}          & \multicolumn{1}{l|}{9.7}         & \blue{70.1}          
\\ \cmidrule(l){1-21} 
SmFM 
& \multicolumn{1}{l|}{2.2}          & \multicolumn{1}{l|}{14.7}          & \multicolumn{1}{l|}{\blue{2.8}}          & \multicolumn{1}{l|}{75.5}          & \multicolumn{1}{l|}{3.8}          & \multicolumn{1}{l|}{19.7}          & \multicolumn{1}{l|}{\blue{3.0}}          & \multicolumn{1}{l|}{69.9}          & \multicolumn{1}{l|}{6.1}          & \multicolumn{1}{l|}{20.7}         & \multicolumn{1}{l|}{\textbf{3.1}} & 65.2          & \multicolumn{1}{l|}{6.1}          & \multicolumn{1}{l|}{12.0}          & \multicolumn{1}{l|}{4.2}          & \multicolumn{1}{l|}{61.1}          & \multicolumn{1}{l|}{7.5}          & \multicolumn{1}{l|}{21.2}         & \multicolumn{1}{l|}{\textbf{4.0}} & 59.7          \\ \cmidrule(l){1-21} 
\textbf{Ours}   
& \multicolumn{1}{l|}{\textbf{1.5}} & \multicolumn{1}{l|}{\textbf{4.1}} & \multicolumn{1}{l|}{\textbf{2.7}} & \multicolumn{1}{l|}{\textbf{83.7}}          & \multicolumn{1}{l|}{\textbf{1.9}} & \multicolumn{1}{l|}{\textbf{1.8}} & \multicolumn{1}{l|}{\textbf{2.8}} & \multicolumn{1}{l|}{\textbf{81.3}}          & \multicolumn{1}{l|}{\textbf{4.8}} & \multicolumn{1}{l|}{\textbf{5.9}} & \multicolumn{1}{l|}{\blue{3.3}} & \textbf{75.8} & \multicolumn{1}{l|}{\textbf{4.1}} & \multicolumn{1}{l|}{\blue{10.2}} & \multicolumn{1}{l|}{\textbf{3.5}} & \multicolumn{1}{l|}{\textbf{71.5}} & \multicolumn{1}{l|}{\textbf{6.1}}          & \multicolumn{1}{l|}{\textbf{6.4}} & \multicolumn{1}{l|}{\textbf{4.0}}          & \textbf{71.1} \\ \bottomrule
\end{tabular}%
}
\caption{Summary of quantitative comparisons for map refinement with different baselines across various datasets and averaged over two different initializations: CDFM~\cite{sun2023spatially} and ULRSSM~\cite{ULRSSM}. The best scores are highlighted in bold, and the second-best scores are highlighted in blue.}
\label{tab:QuantRefinement}
\end{table*}

\subsection{Implementation details}
\label{subsec:ImplDetails}
All our supervised training is performed over the first 50 shapes from SCAPE~\cite{Anguelov2005SCAPESC} dataset and the first 10 shapes from the FAUST dataset~\cite{Bogo:CVPR:2014}. Our choice stems from the compactness of the dataset size and the range of deformation between each pair. We use a 4-layered MLP of 256 feature dimensions with the spectral projection followed by 2 layers without spectral projection with 9-dimension output, which is our Jacobian. We train with ADAM optimizer for a total of 50 epochs, with an initial learning rate of 1e-3 decayed progressively to 1e-5. The coefficients across different objective functions used in this paper are respectively $\alpha_1=1$, $\alpha_2=10.$, $\alpha_3=2$, $\alpha_4=20000$, $\alpha_5=150000$, $\alpha_6=20.$ and $\alpha_7=10.$ respectively. We used 128 eigenfunctions for smoothing the learned feature while we used 40 eigenfunctions to project the embedding of the shape. While training our network, as form of data augmentation, we vary the number of eigenfunctions used to construct the spectrally smooth embedding. Our full implementation and data will be released upon publication. 

\section{Experimental Results}
In this section, we demonstrate the utility of our deformation module, LJN, across three tasks - namely, Map Refinement in Section~\ref{subsec:Refinement}, Unsupervised Deformation and Mapping in Section~\ref{subsec:Unsupervised} and finally Interactive Editing in Section~\ref{subsec:Editing}. Before delving into individual tasks, we first elaborate the metrics used to gauge different methods in Subsection~\ref{subsec:Metrics}. For supervised experiments, our network is trained on 60 shape pairs from FAUST~\cite{Bogo:CVPR:2014} and SCAPE~\cite{Anguelov2005SCAPESC} datasets. 

\subsection{Evaluation Metrics}
\label{subsec:Metrics}
For evaluating shape correspondence, we used the standard mean geodesic error~\cite{Kim2011BIM}. Additionally, we assessed map inversion, the Dirichlet energy~\cite{magnet2022smooth} of the map, and coverage~\cite{Ren2018}. For completeness, we provide elaboration on these metrics below.

\begin{enumerate}
    \item Map Inversion: This metric measures the change in orientation induced by the map. For surface correspondence techniques, matching via nearest neighbors search can result in a $180^{\circ}$ twist, commonly referred to as the candy-wrapper effect (see \cite{abulnaga2022symmetric} Fig. 2). To quantify this, we use map-inversion metric, which penalizes the deviation in the sign of the dot product between the normal on the target shape and the mapped shape. The mapped shape normal is estimated by \emph{permuting} the target vertices via $\Pi_12$ and estimating the normals using the orientation prescribed by the \emph{source mesh}. More precisely, let $\mathcal{S}_1: \{\mathcal{V}_1, \mathcal{F}_1\}$ and $\mathcal{S}_2: \{\mathcal{V}_2, \mathcal{F}_2\}$ be the source and the target mesh. We let $\vec{\hat{n}}^i_1$ be the normal corresponding to the mesh $\hat{\mathcal{S}}_1: \{\Pi_{12} \mathcal{V}_2, \mathcal{F}_1\}$, whose vertices image of $\mathcal{V}_2$ under $\Pi_{12}$ with winding defined by $\mathcal{F}_1$. Since $\Pi_{12}$ acts as a map between vertices, let $\vec{\hat{n}}^j_2$ be the normal at vertex index $j$ on $\mathcal{S}_2$, such that $\Pi_{12} [i,j] = 1$. Note $\vec{\hat{n}}^j_2$ is defined by the orientation of $\mathcal{S}_2$. Then, map inversion is quantified as 

\begin{equation}
    \mathrm{Inv} := \frac{ \sum\limits_{j=1}^{\| \mathcal{V}_2 \|} \mathbf{1}\left( \sum\limits_{i=1}^{m} \vec{\hat{n}}^i_1 \cdot \vec{\hat{n}}^j_2 < 0 \right)}{\| \mathcal{V}_2 \|} \times 100
    \label{eqn:Inversion}
\end{equation}

    \item Dirichlet Energy: Following the standard convention used in previous works~\cite{RHM,magnet2022smooth}, we define map smoothness as $\| \Pi_{12} \mathcal{V}_2 \|_{\Delta_1}$. For uniformity, we scale all shapes to have a unit area before computing this energy.

    \item Coverage: Similar to ~\cite{Ren2018}, we define coverage as the percentage sum of the unique area of the vertex image given by $\Pi_{12} \mathcal{V}_2$. Note that since we scale shapes to have a unit area, normalization is not required.

\end{enumerate}

\subsection{Map-Refinement}
\label{subsec:Refinement}

\subsubsection{Overview}
Given an approximate shape correspondence, either as a fuzzy point-wise map or a functional map~\cite{ovsjanikov2012functional}, our task is to produce a refined point-wise map. To that end, we first construct the spectrally projected input signal (c.f Eqn~\ref{eqn:BlurInference}), then, perform a feedforward pass over LJN to obtain the Jacobian, from which we compute the embedding using Eqn~\ref{eqn:EmbedRecoveryBacksub}. Finally, the refined point-wise map is obtained via Eqn~\ref{eqn:FinalP2P}.

\subsubsection{Setup}
  As our input correspondence, we use two recent Deep Functional Map methods~\cite{ULRSSM,sun2023spatially} as our baseline maps, for which open-source implementations are available. We evaluated our approach and all baselines on the FAUST, SCAPE~\cite{Ren2018} and the SHREC-19 dataset~\cite{donati2020deepGeoMaps} for near-isometric shape correspondence. We used the inter-class category of Deforming-4D~\cite{magnet2022smooth} and SMAL-remeshed~\cite{li2022attentivefmaps} for non-isometric correspondence. We use geodesic error~\cite{Kim2011BIM} (cm), map coverage~\cite{Ren2018} (\%), map inversions (\%) and map smoothness as evaluation metrics. 
\subsubsection{Comparisons}
We compare our method against axiomatic \emph{iterative} map-refinement methods: ZoomOut (ZO) ~\cite{MelziZO} , Discrete Optimization (DZO)~\cite{Ren2021}, and Smooth Functional Maps (SmFM) ~\cite{magnet2022smooth}.  More details on the evaluation and implementation of baselines are provided in the Supplementary. Additionally, we consider two more plausible alternatives. First, we set \(\alpha_5=0\), so that the recovered embedding (viz., the refined map) does not rely on the learned Jacobians. Secondly, we use the coordinate function `pulled' by Functional Map (cf. Equation~\ref{eqn:BlurInference}) to recover the map. i.e, we solve for the map as $\Pi_{12} = \mathrm{NNSearch} (\bar{\mathcal{V}}_1, \mathcal{V}_2)$ (c.f Eqn~\ref{eqn:FinalP2P}). We denote this baseline as ``Spec Proj''.

\begin{figure}[t]
  \centering
  \includegraphics[width=\linewidth]{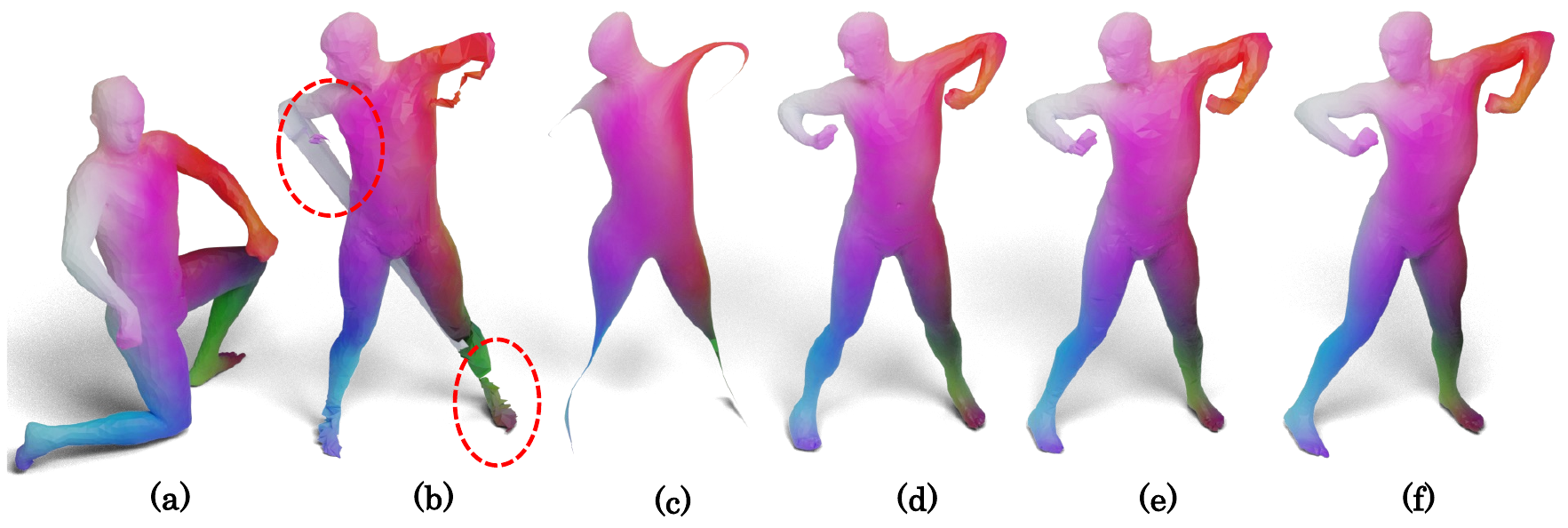}
  \caption{Illustrating \emph{how} LJN achieves high map-refinement accuracy: Given (a) the source shape and (b) an input map containing high-frequency artifacts, (c) we convert it to the Functional Map and \emph{pull} the coordinates. The pulled coordinates are devoid of high-frequency artifacts. Subsequently, (d) our reconstruction and (e) the obtained map are detail-preserving. (f) denotes the target shape. We color-code correspondence from refined maps.}
\label{fig:How}
\end{figure}

\begin{figure}[t]
  \centering
  \includegraphics[width=\linewidth]{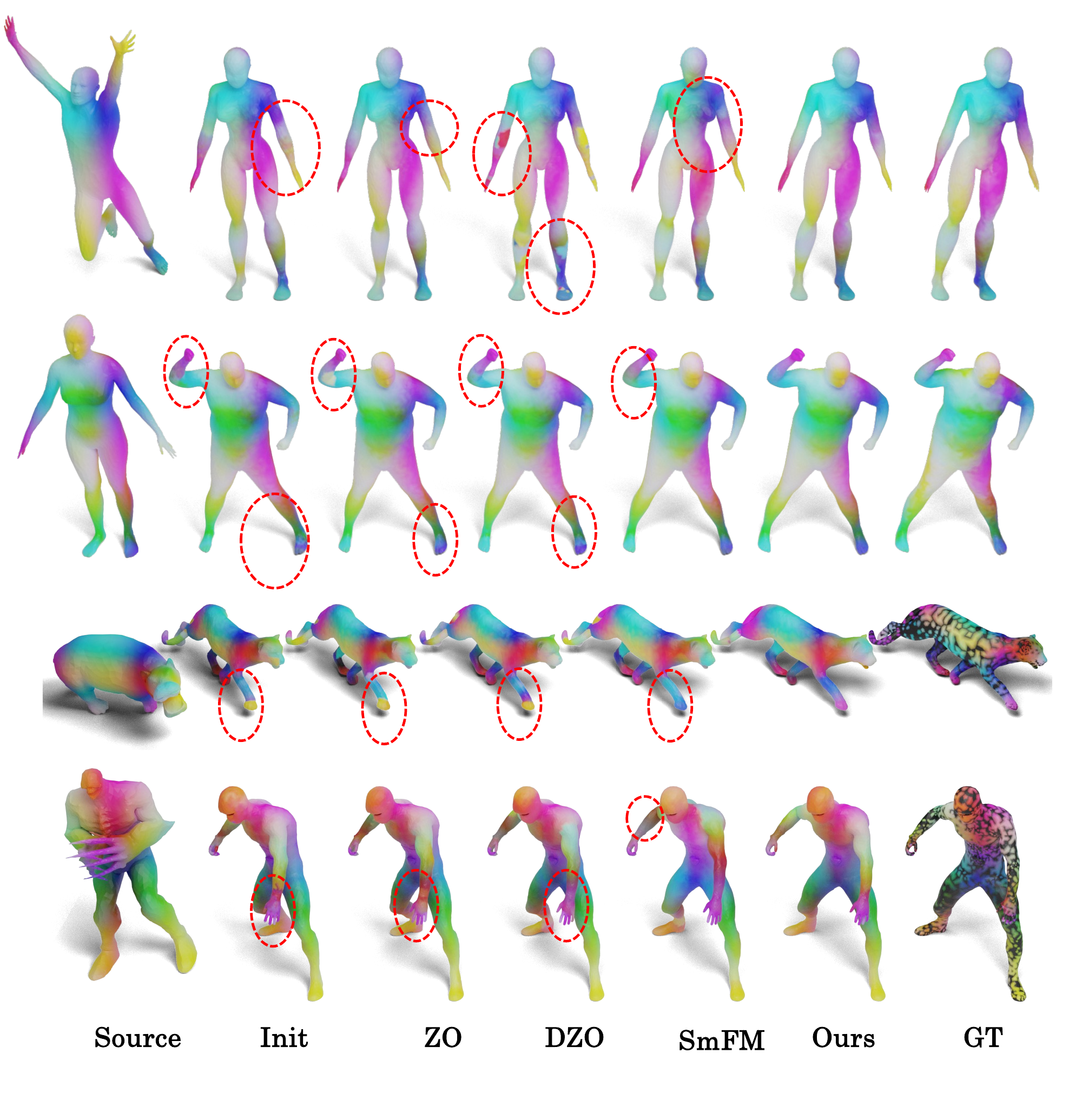}
  \caption{Qualitative results for map refinement. The first two rows depict near-isometric shape pairs from SHREC'19~\cite{SHREC19} dataset, while the last two rows show non-isometric datasets: SMAL~\cite{donati2020deepGeoMaps} and Deforming4D~\cite{magnet2022smooth}. We note that LJN not only preserves correspondence but also improves the smoothness of the produced map. For initialization, we used CDFM~\cite{sun2023spatially} for the first two rows and ULRSSM~\cite{ULRSSM} for the latter two.}
  \label{fig:QualRefinement}
\end{figure}

\subsubsection{Discussion}
\begin{figure}[t]
  \centering
  \includegraphics[width=0.91\linewidth]{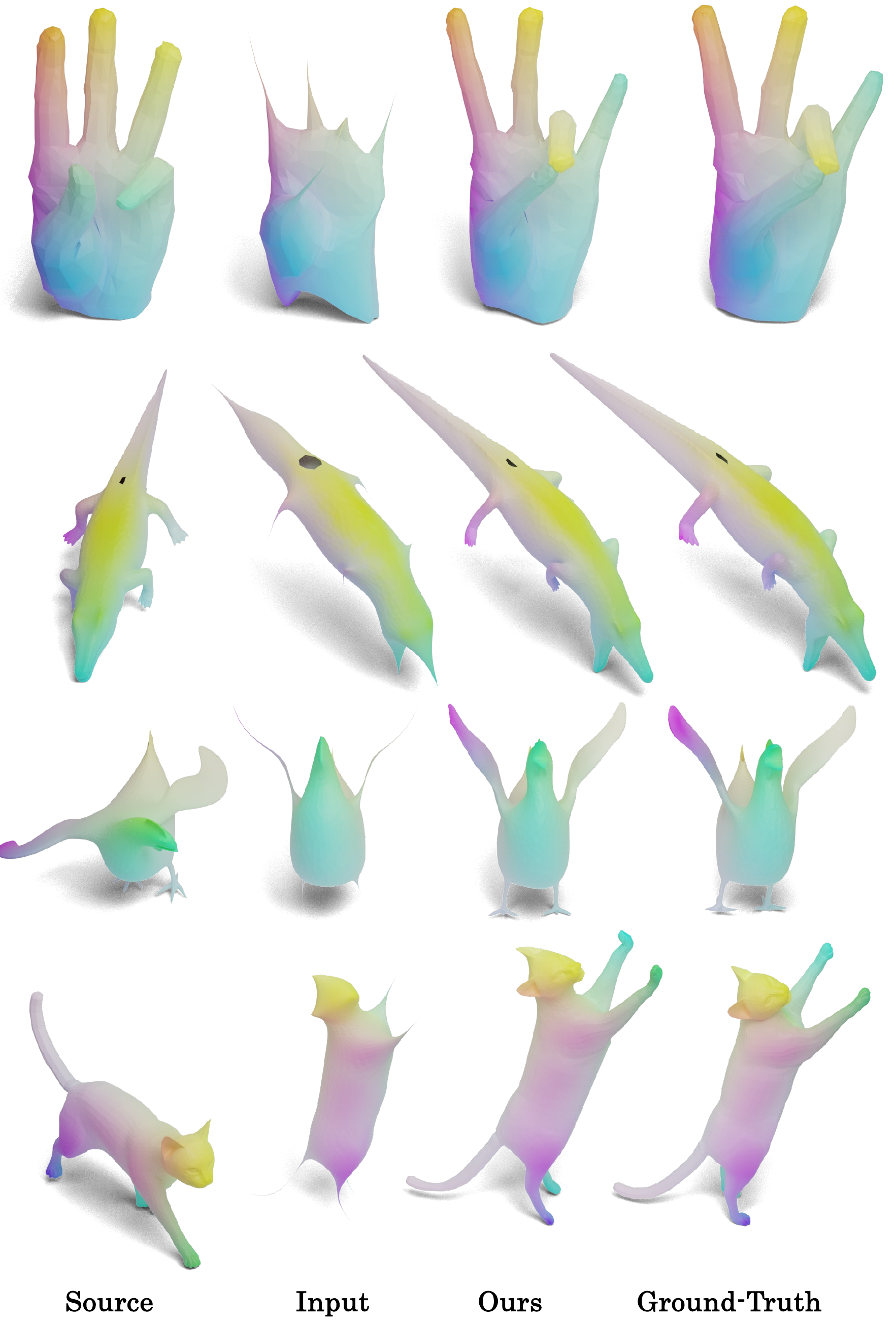}
  \caption{Generalization of LJN to different object categories. Trained on 60 pairs of human shapes, our LJN shows robust generalization to unseen categories with near-perfect reconstruction. Meshes in the first row are from the MANO dataset~\cite{MANOSIGGRAPHASIA2017}, the subsequent three rows are from the Deforming-4D Dataset~\cite{magnet2022smooth}, and the last row is from the TOSCA dataset~\cite{Ren2018}. Shapes in rows 2-5 do not share identical triangulations. To estimate the deformation, we used a $40 \times 40$ FMap. The embedding is then recovered following Section~\ref{subsec:Inference}.}
  \label{fig:DeformationGeneralization}
\end{figure}

The quantitative evaluation averaged across two initializations, is summarized in Table~\ref{tab:QuantRefinement}. Our approach consistently outperforms iterative baseline methods ZO, DZO, and SmFM across various benchmarks and evaluation metrics. Notably, despite being trained solely on human shapes, our method demonstrates remarkable generalization in modeling animal deformations from the SMAL dataset. Additionally, maps recovered by our method exhibit less distortion compared to those from SmFM \cite{magnet2022smooth}, which \textit{explicitly} aims to minimize Dirichlet energy. This improvement is attributed to the truncation of the basis inherent in their approach, whereas our deformation operates in the spatial domain ($\mathbb{R}^{3}$). We illustrate the effects of map smoothness and overall map refinement accuracy in Figure~\ref{fig:QualRefinement}. Furthermore, we show qualitative reconstruction across more object categories with different triangulations in Figure~\ref{fig:DeformationGeneralization}. Finally, we provide an intuitive explanation in Figure~\ref{fig:How} to illustrate \emph{how} our approach achieves superior performance. Overall, LJN shows state-of-the-art results, which is remarkable due to the difficulty of the problem and the presence of strong recent baselines.

\subsection{Unsupervised Deformation and Mapping}
\label{subsec:Unsupervised}
\begin{figure}[t]
  \centering
  \includegraphics[width=0.91\linewidth]{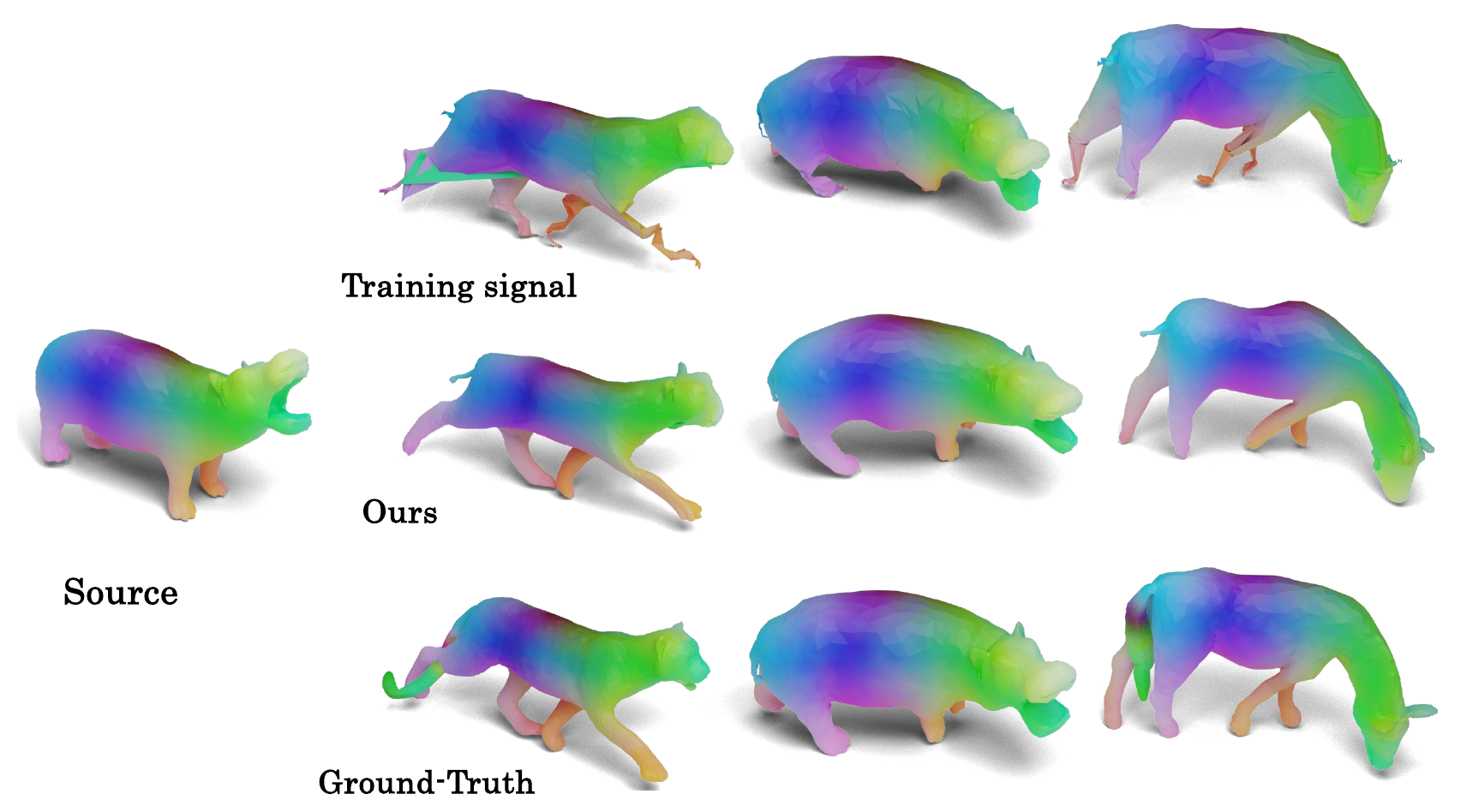}
  \vspace{-0.3cm}
  \caption{Qualitative reconstruction results of training LJN without supervision. For a given source, the first row depicts the training signal used in our unsupervised loss (c.f Eqn~\ref{eqn:UnsupLoss}). In spite of training with noisy supervision, LJN recovers cleaner deformation as depicted in the subsequent row. Please refer to Tab.3 in Appendix for quantitative result.}
\label{fig:UnsupDef}
\end{figure}

\begin{figure}[t]
  \includegraphics[width=\linewidth]{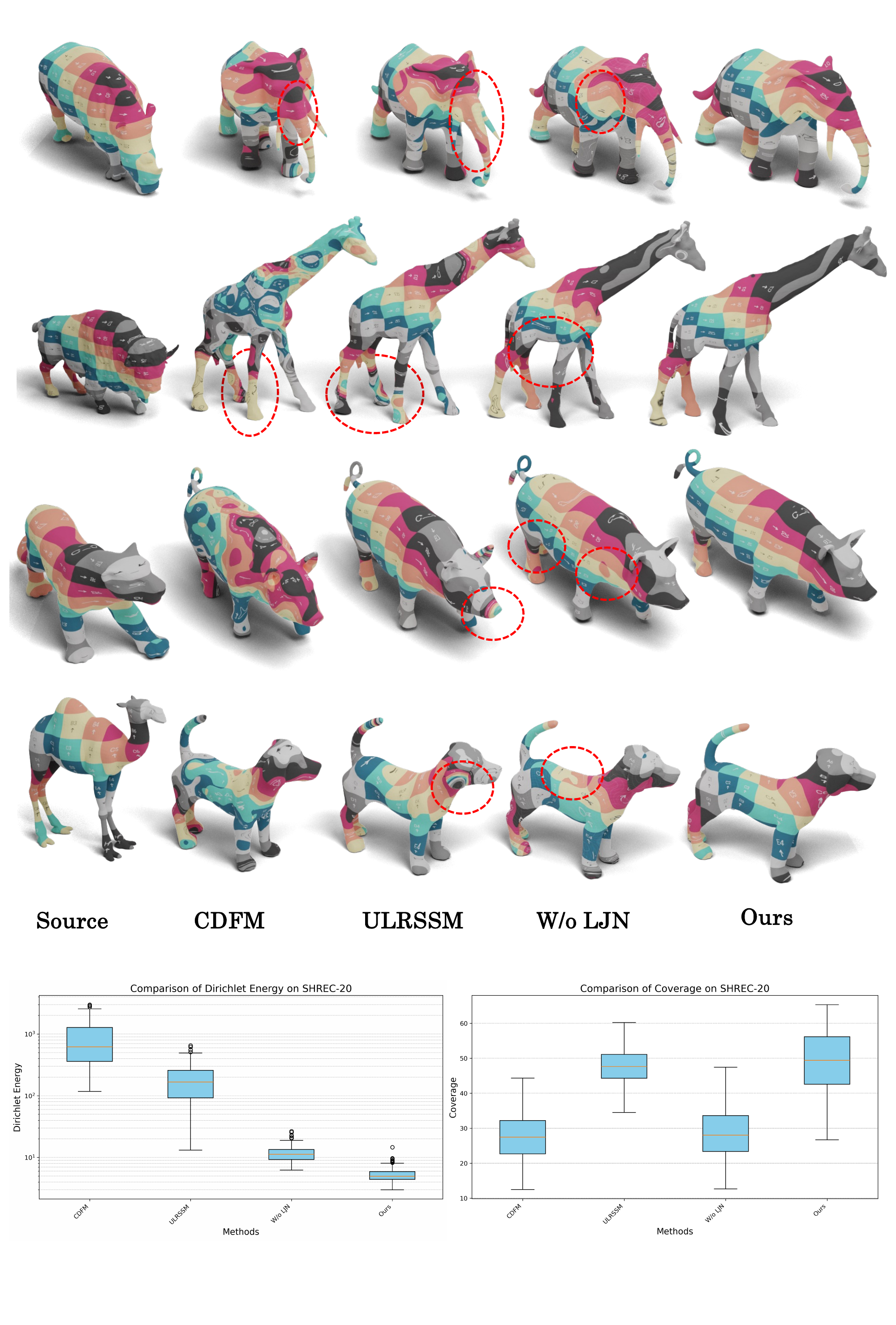}
  \vspace{-1.2cm}
  \caption{Qualitative and Quantitative results for unsupervised shape correspondence on the SHREC'20~\cite{DykeShrec20} dataset. We compare our results with those from CDFM~\cite{sun2023spatially}, ULRSSM~\cite{ULRSSM}, and a variant of our approach (W/o LJN) where the point-wise map is obtained by following the standard procedure (cf. Eqn~\ref{eqn:FM2P2P}) instead of our proposed method (cf. Eqn~\ref{eqn:FinalP2P}). In the last column, we show the point-wise map recovered by LJN. ULRSSM produces plausible maps but exhibits severe distortion in various highlighted regions. In contrast, our approach preserves correspondence while minimizing distortion.}
  \label{fig:UnsupMapping}
\end{figure}

Given a collection of shapes with different triangulations, we aim to produce the deformation and the map as detailed in Section~\ref{subsec:unsupervised}. We use the SMAL~\cite{li2022attentivefmaps} and SHREC20~\cite{DykeShrec20} datasets for two separate experiments, which include a variety of largely non-isometric animal shapes. For the former setup, we train on 28 shape pairs in SMAL across 5 categories and evaluate on 20 shape pairs spanning 3 unseen categories. In Figure~\ref{fig:UnsupDef}, we provide the qualitative deformations obtained from unsupervised training on SMAL. Despite being supervised with a noisy training signal (first-row), LJN produces detail-preserving deformation. Due to the limited training data in SHREC20, following~\cite{ULRSSM}, we perform zero-shot learning on each pair. Owing to the absence of dense correspondence in the dataset, we only provide coverage and smoothness as the quantitative measures. We compare with recent DFM methods ULRSSM~\cite{ULRSSM}, CDFM~\cite{sun2023spatially}, and a variant of our approach where point-wise map is extracted without our deformation module (c.f Eqn~\ref{eqn:FM2P2P}), denoted as W/o LJN in Figure~\ref{fig:UnsupMapping}.

\paragraph{FAUST-Challenge}
To demonstrate the generalizability of our approach to real-world scans, we evaluated our \textit{unsupervised} LJN (Sec. \ref{subsec:unsupervised}) on the FAUST-Challenge dataset~\cite{Bogo:CVPR:2014}. This dataset contains 100 shape pairs across near-isometric (INTRA) and non-isometric (INTER) registration challenges. The scans include various artifacts such as non-manifold edges, self-intersections, and topological noise. Despite not having specific mechanisms for handling data imperfections and without hyperparameter tuning, LJN achieves comparable performance to unsupervised state-of-the-art methods, with a mean error of \textbf{3.89} cm for INTRA and \textbf{2.62} cm for INTER challenges respectively. We remark especially on the low error on the more challenging INTER dataset. For the INTER dataset, our average error is dominated by three outlier pairs with significant error due, especially to symmetry mixing, which have an average error of 30 cm. We believe that such errors could be potentially be reduced by using symmetry disambiguation \cite{donati2022complex,donati2022DeepCFMaps} and post-processing \cite{vestner2017product,MelziZO,huang2020consistent} techniques. 

\subsection{Interactive Editing}
\label{subsec:Editing}
\begin{figure}[t]
  \centering
  \includegraphics[width=\linewidth]{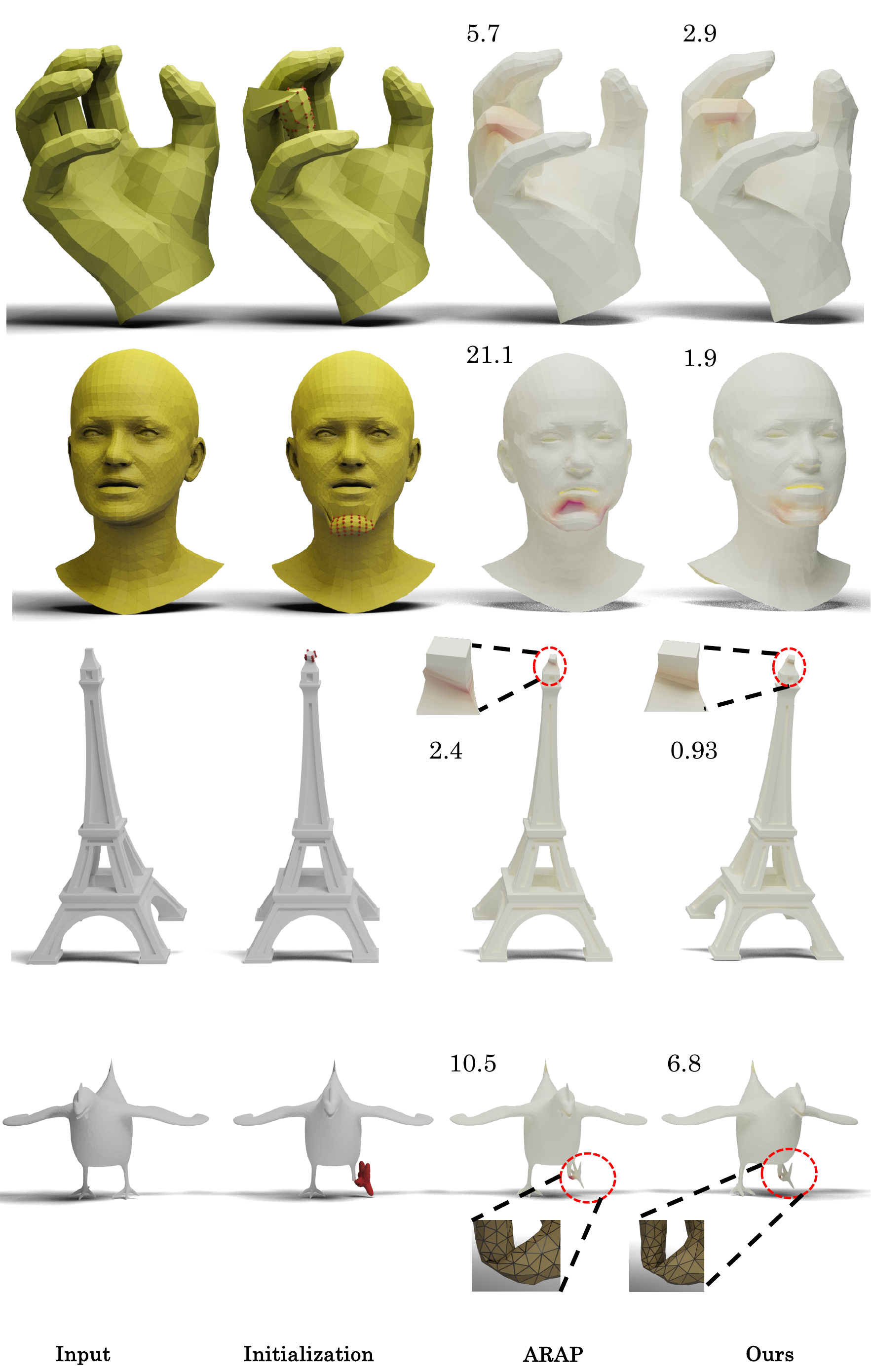}
  \caption{Qualitative results for handle-based deformation. We selected a set of handles (shown in red) and applied a deformation by displacing those vertices. We then compare the reconstruction between our approach and ARAP. The error plot on the outputs depicts the distribution of symmetric Dirichlet energy, with the average value given in the annotation. LJN produces a more plausible mesh with smooth embedding compared to ARAP~\cite{Sorkine2006}. Please refer to the supplementary material for more quantitative insights. }
  \label{fig:Handle}
\end{figure}

For this task, we re-trained our network on the same dataset of 60 human shape pairs, but with the closest rotation as the input signal to LJN. We evaluated two setups for handle-based deformation. In the first, we performed handle-based manual deformation, where a user drags a selected portion of the mesh. This was done by selecting regions of the mesh in Blender and applying a sequence of rigid transformations to the selected vertices. We evaluated our method on five different object categories, as shown in Figure~\ref{fig:Handle}. We compared our approach with ARAP and observed that our method produces lower distortion, measured by symmetric Dirichlet~\cite{Smith2015}. LJN produces a smoother embedding than ARAP without undesirable artifacts such as changes in pose (row 2) and self-intersection (row 4).

\begin{figure}[t]
  \includegraphics[width=\columnwidth]{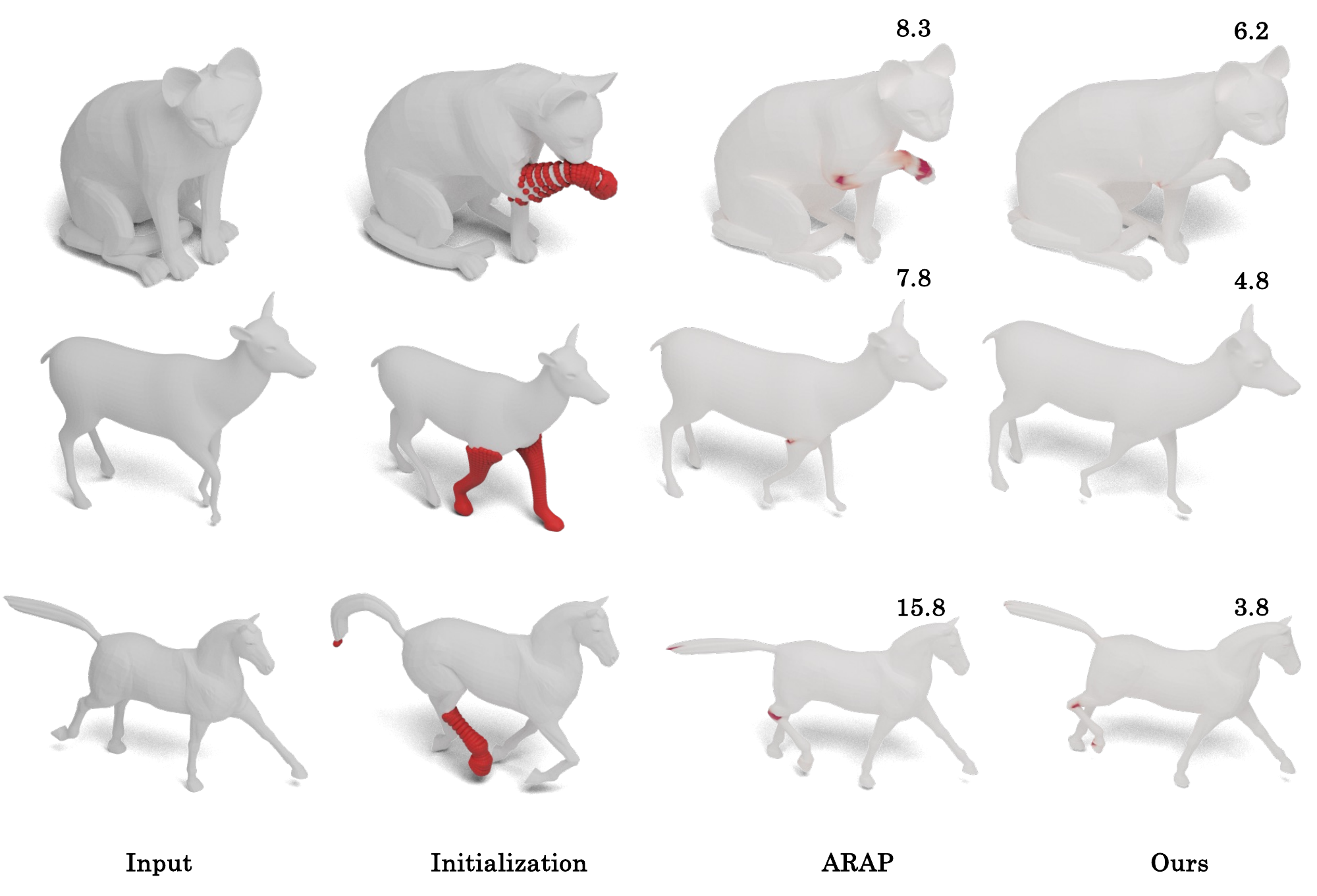}
  \caption{We perform interactive editing on animation sequence meshes by selecting handles as vertices with the largest motion between the first and second columns. Our approach produces more plausible meshes in bending regions compared to ARAP. The values annotated above are symmetric Dirichlet scaled by $10^{-2}$}
  \label{fig:LargeDefo}
\end{figure}

In the second setup, we used datasets containing meshes corresponding to deformation sequences. Please note that this second experimental setup is more challenging since the deformation is significantly larger. We considered animals from the ~\cite{Sumner,li20214dcomplete} datasets and automatically selected key points, corresponding to the group of vertices that undergo the largest displacement. Given the new positions of the key points, we estimated the Jacobians between the source mesh and the suggested deformation, projected them to the closest rotation, and estimated the smoother deformation predicted by LJN. We evaluated this on three animals—Cat, Buck, and Horse—as shown in Figure~\ref{fig:LargeDefo}. LJN produces more plausible deformation, especially at regions pertaining to bending in comparison with ARAP. 

\begin{figure}[t]
  \includegraphics[width=\columnwidth]{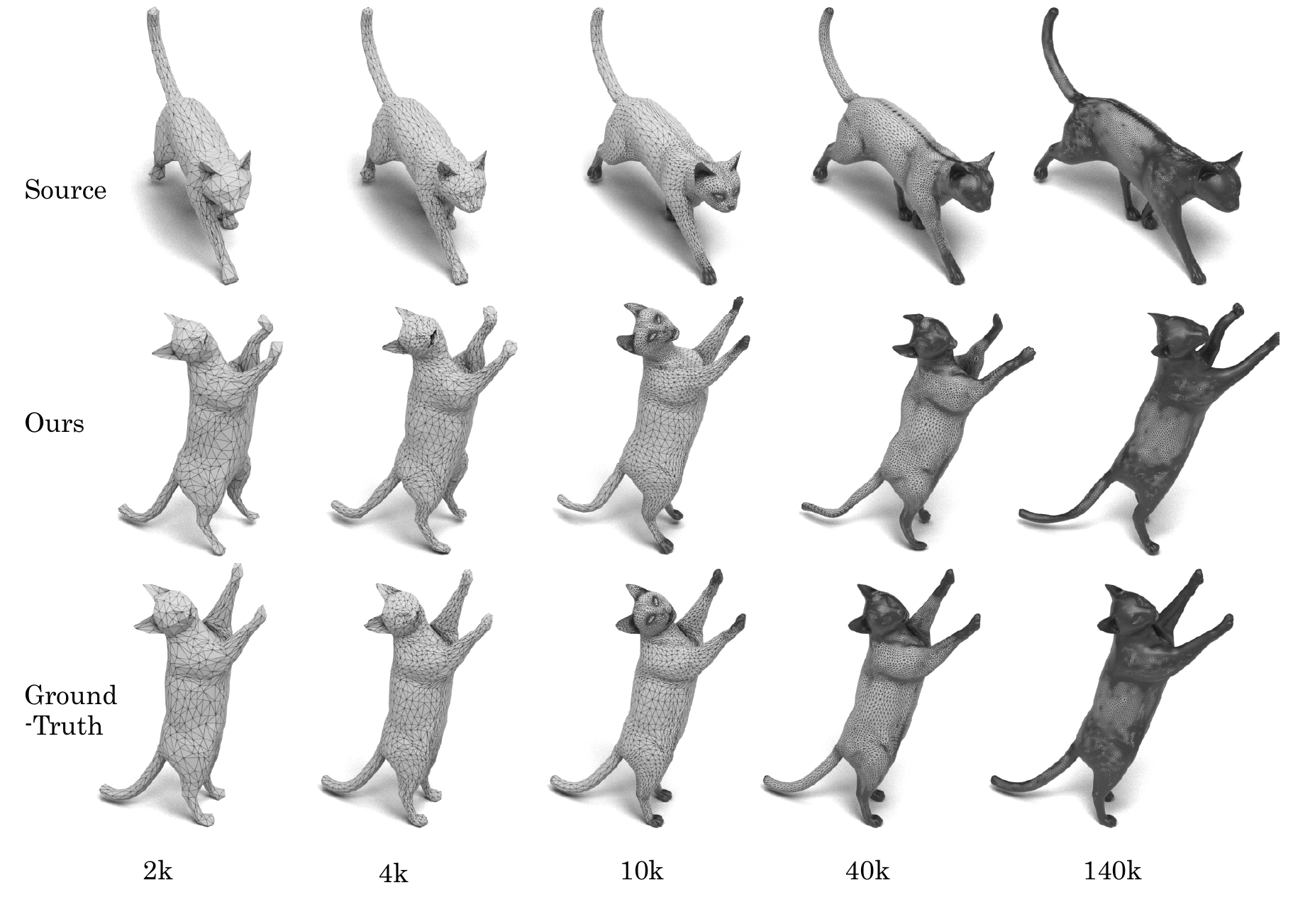}
  \caption{Comparison of deformation between the same source and target geometry across different tessellations. The face counts are annotated along the rows. Across varying resolutions, our deformation remains faithful to the target geometry.}
  \label{fig:Resolution}
\end{figure}

\section{Additional Experimental Insights }
\label{subsec:AddExp}
We first perform an ablation study in Section~\ref{subsec:Ablation}, where we empirically justify the simple architecture of LJN. Next, we provide a runtime comparison between our approach and different baselines across various datasets evaluated in Section~\ref{subsec:RunTime}.

\begin{table}[]
\centering
\resizebox{0.7\columnwidth}{!}{%
\begin{tabular}{@{}|l|ll|ll|@{}}
\toprule
Module          & \multicolumn{2}{l|}{FAUST}      & \multicolumn{2}{l|}{SCAPE}       \\ \midrule
                & \multicolumn{1}{l|}{CD}  & Geod & \multicolumn{1}{l|}{CD}   & Geod \\ \midrule
$\alpha_1=0$    & \multicolumn{1}{l|}{5.1} & 1.7  & \multicolumn{1}{l|}{10.0} & 2.4  \\ \midrule
Displacements   & \multicolumn{1}{l|}{5.2} & 1.8  & \multicolumn{1}{l|}{10.4} & 2.3  \\ \midrule
Face-based      & \multicolumn{1}{l|}{7.4} & 2.1  & \multicolumn{1}{l|}{13.1} & 3.0  \\ \midrule
W/o Smooth      & \multicolumn{1}{l|}{4.8} & 1.8  & \multicolumn{1}{l|}{10.7} & 2.6  \\ \midrule
Baseline (Ours) & \multicolumn{1}{l|}{3.9} & 1.5  & \multicolumn{1}{l|}{7.6}  & 2.1  \\ \bottomrule
\end{tabular}%
}
\caption{Ablation study on map refinement and deformation reconstruction on the test set of the FAUST and SCAPE datasets~\cite{Ren2018}. CD denotes Chamfer's Distance and Geod is the mean geodesic error~\cite{Kim2011BIM}}
\label{tab:Ablation}
\end{table}

\subsection{Ablation Studies}
\label{subsec:Ablation}
\begin{figure}[t]
  \centering
  \includegraphics[width=0.9\linewidth]{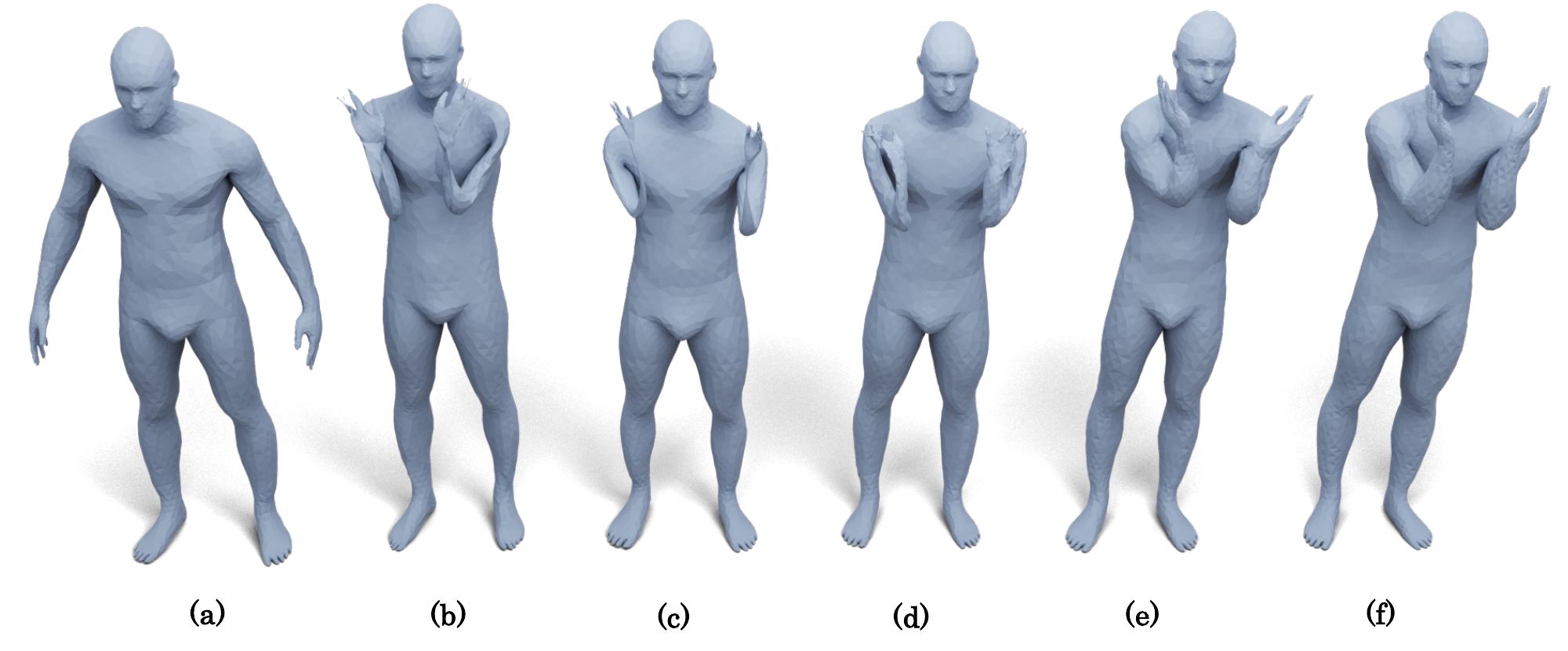}
  \caption{Ablation study performing a qualitative comparison between possible alternatives. Given the (a) Source shape, (b) is the reconstruction from displacement field, (c) denotes Face-based discretization, and (d) is the reconstruction without using the feature smoothness in the MLP. Our reconstruction is depicted in (e) and finally, (f) denotes the target shape. }
  \label{fig:Ablation}
\end{figure}

We perform an ablation study to understand the effect of supervising Jacobians, using displacement fields instead of Jacobians, input discretization (face vs. vertex), and impact of feature smoothing (Projection of features to the eigenspace of LBO Operator). To this end, we compare the reconstruction accuracy of the deformation and the correspondence error by measuring the mean geodesic discrepancy. We summarize our results in Table~\ref{tab:Ablation}. Face-based discretization lacks information sharing within the local neighborhood, resulting in implausible deformations leading to poor reconstruction results. On the other hand, LJN promotes smoother feature space and as a result, achieves more plausible deformation. We visualize the problems inherent to alternatives we tried in Figure~\ref{fig:Ablation}. All baselines produce artifacts in reconstruction corresponding to high-frequency parts of the shape while ours recovers the near-exact geometry. For a fair comparison, all methods were trained on the same training data, following the same hyper-parameters.

\begin{figure}[t]
  \centering
  \includegraphics[width=0.9\linewidth]{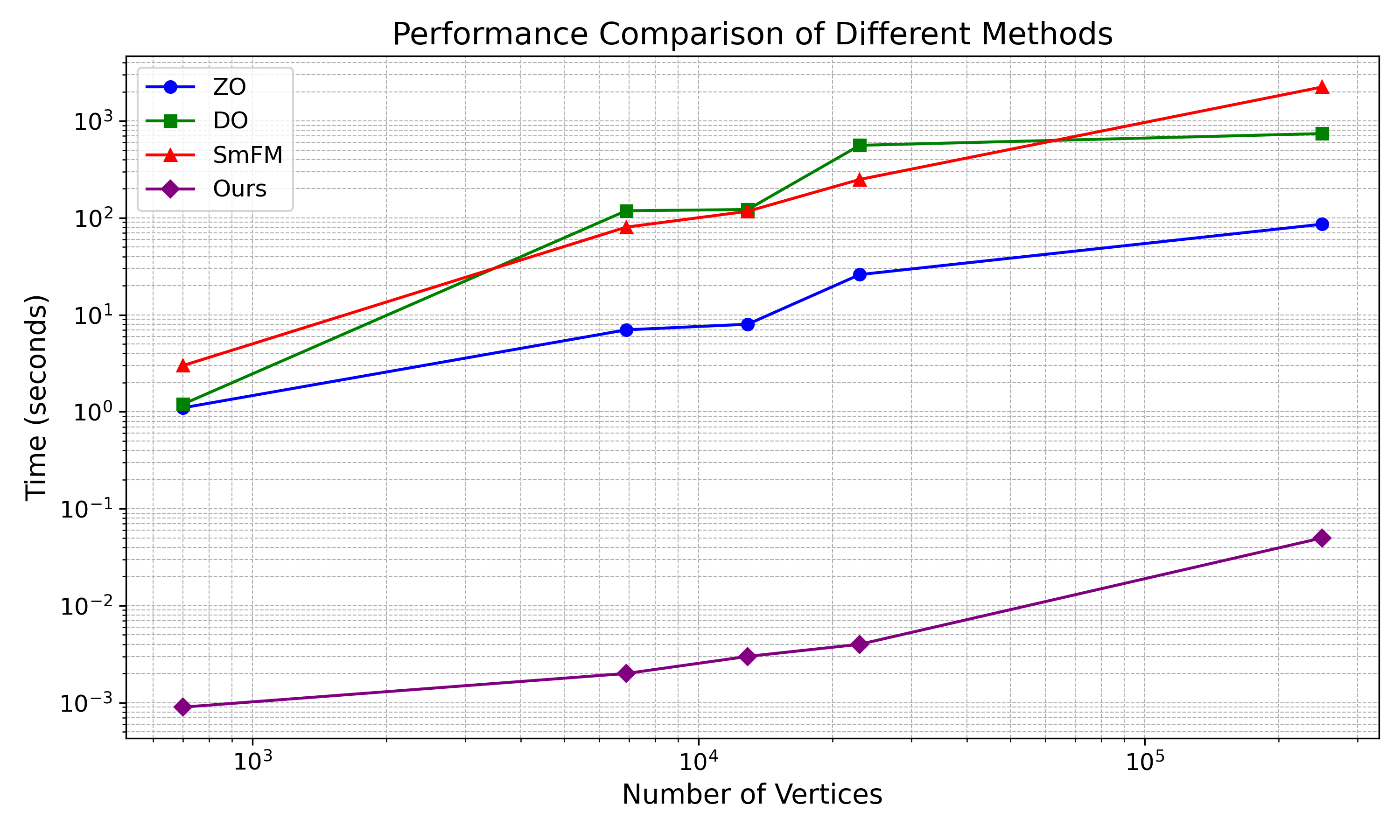}
  \caption{Performance (run-time) comparison of different methods. Our LJN shows orders of magnitude faster performance than other baselines. }
  \label{fig:Performance}
\end{figure}

\subsection{Robustness to changes in tesselation}
We analyze the effect of tesselation on LJN. To this end, we take a pair of shapes from the animal dataset in~\cite{Sumner} and re-mesh to generate shapes with varying triangulation. Namely, we generate the same shape with varying triangulation ranging from 2K faces to 200K faces. We apply mesh decimation with Quadric Edge collapse to obtain the simplified meshes with reduced vertex and face count. To increase the resolution, we apply loop subdivision until a mesh with the desired face count is obtained. This re-meshing is done to the pair of initial meshes. Then, we compute the spectrally projected Jacobians between (c.f. Eqn~\ref{eqn:SpectralBlur}) and estimate the deformation from the network prediction. We compare different mesh resolutions in Figure~\ref{fig:Resolution}. Across different discretizations, our LJN produces a deformation which is consistently faithful to the target geometry.

\begin{figure*}
  \centering
  \includegraphics[width=\linewidth]{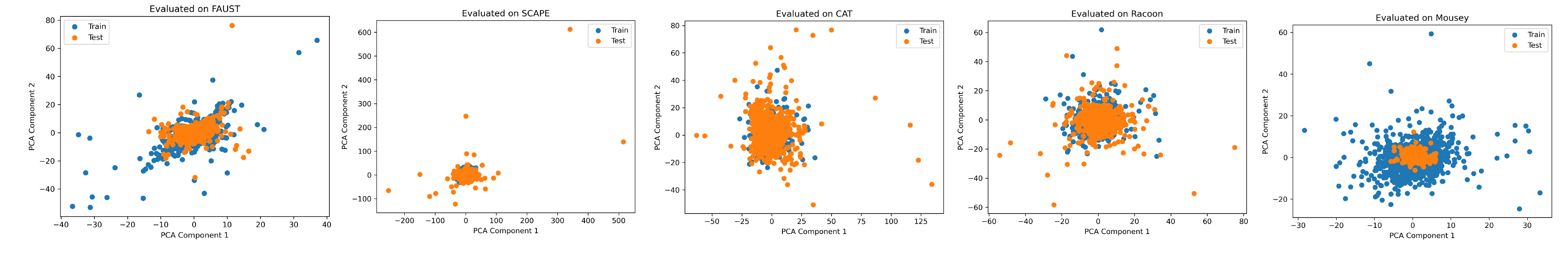}
  \caption{We compare the input signal between different evaluation datasets and our \emph{standard} training set, which was identical across two supervised experiments detailed in the main paper using Principal Component Analysis (PCA). The evaluation datasets we compare with respectively are FAUST~\cite{Bogo:CVPR:2014}, SCAPE~\cite{Anguelov2005SCAPESC}, sequences of Cats from ~\cite{Sumner}. Racoon and Mousey are two object categories from the Deforming4D Dataset~\cite{li20214dcomplete}}
  \label{Fig:PCA}
\end{figure*}

\subsection{Baseline Details}
\label{subsec:Baseline}
Across all experiments, we used the open-source code released by the authors of the respective papers for baseline comparisons. Starting with axiomatic refinement methods, we ran ZoomOut beginning with a $20\times20$ Functional Map, upsampling with a step-size of 5 to reach the resolution of a $120\times120$ sized functional map. For SmFM~\cite{magnet2022smooth} and Discrete Optimization~\cite{Ren2021}, we used the Python implementation available in the \href{https://github.com/RobinMagnet}{PyFM library}. Specifically, for Discrete Optimization, we initialized the Functional Map at $20\times20$ and upsampled it to $100\times100$, whereas for SmFM, we initialized a $10\times10$ Functional Map and upsampled it to $70\times70$. For Consistent FM~\cite{sun2023spatially} and ULRSSM~\cite{ULRSSM}, we used the author-provided code and pre-trained models. For Consistent-FM, we used the point-wise map corresponding to $80\times80$ Functional Map. Finally, for handle-based deformation, we used the libigl implementation of ARAP~\cite{libigl} as our baseline. 

\subsection{Run-Time Comparison}
\label{subsec:RunTime}
We perform a run-time comparison between our approach and different baselines for the task of map refinement. We show this comparison in Figure~\ref{fig:Performance}, evaluated over shape pairs consisting of different number of vertices. We do not count the pre-processing time since both LJN and all baselines require the Eigen decomposition of the LBO operator. To recall, our LJN performs map refinement in a single feedforward pass followed by back substitution, while the baselines are iterative refinement techniques. As a result, LJN is orders of magnitude faster than baselines. These experiments were performed on a machine with AMD 7302 and Nvidia A100 GPU.

\subsection{Discussion on Generalization}
\label{subsec:PCA}
We provide an empirical overview to explain why our approach generalizes across different categories of objects. To demonstrate this, we compare the input signal from our training set with the input signals across different object categories used in our qualitative and quantitative evaluations. Recall that our input signal is the Jacobian corresponding to the spectrally projected target shape, averaged over vertices. We compare this input signal, defined at vertices, across different datasets using Principal Component Analysis (PCA), as depicted in Figure~\ref{Fig:PCA}. The strong correlation between the input signals in our training and test datasets highlights the remarkable ability of LJN to generalize.

\begin{figure}[t]
  \includegraphics[width=\columnwidth]{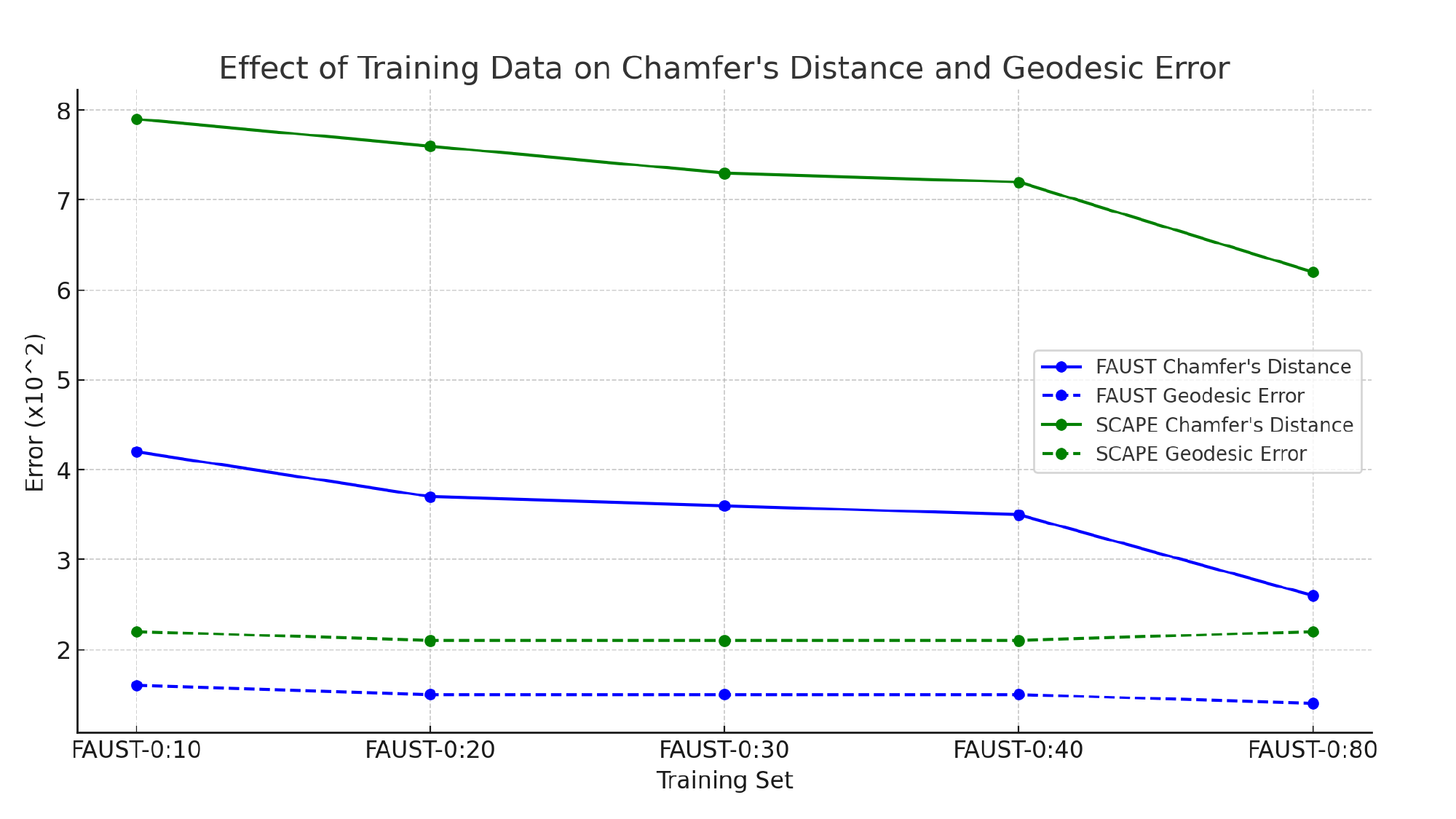}
  \caption{Understanding the effect of training data required to achieve optimal performance. Trained on 10 shapes, our approach demonstrates comparable performance for shape correspondence as a model that has been trained on 80 shapes.}
  \label{fig:TrainingData}
\end{figure}

\subsection{Effect of Training Data}
\label{subsec:TrainingData}
LJN is highly data-efficient, requiring only 60 pairs of human shapes to demonstrate significant generalization across different shape categories. In this section, we empirically investigate the minimum data required to achieve optimal performance. We begin with a dataset of 10 shapes and incrementally increase it to 80 shape pairs, all from the FAUST dataset. We evaluate the performance of LJN over different training set using reconstruction and shape correspondence metrics on both the FAUST and SCAPE datasets, following the map-refinement setup detailed in Section~\ref{subsec:Refinement}. The results, visualized in Figure~\ref{fig:TrainingData}, indicate that LJN achieves accurate results with as few as 10 shape pairs. Moreover, evaluating SCAPE~\cite{Anguelov2005SCAPESC} also yields accurate results despite the pairs being near-isometric and having different connectivity than the training (FAUST) shapes.

\begin{figure}[t]
  \includegraphics[width=\columnwidth]{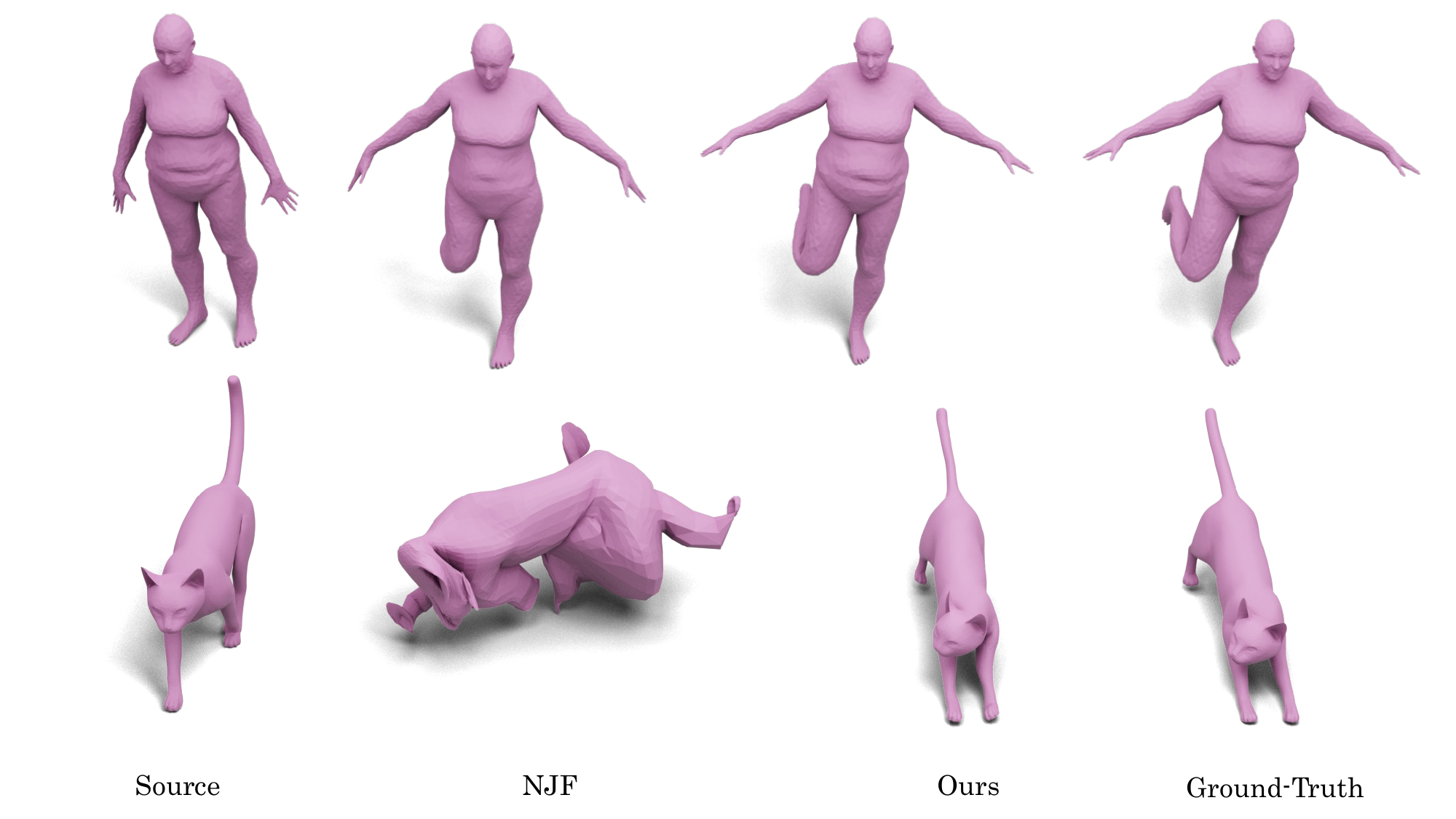}
  \caption{Comparison of reconstructions produced by a global encoding (NJF) and our LJN which uses local signals to learn deformation. NJF and our approach produce near-perfect deformations when evaluated on a pair of shapes from the same training category (first row). However, NJF fails to produce cross-category generalization, where, LJN shows consistent reconstruction (second row).}
  \label{fig:GlobalCoding}
\end{figure}

\subsection{Comparison with Global Encoding}
We also compare our LJN approach with the original Neural Jacobian Fields (NJF) ~\cite{aigerman2022neural} method which relies on a \textit{global} latent code to define a deformation of a shape. As mentioned above, the global nature of the latent code along with the supervised nature of NJF imply that that approach requires extensive data and \textit{per-category training} to produce plausible deformations. Thus, the basic version of NJF cannot be directly compared with the various experimental setups in this paper, as its use of global encoding fundamentally limits its generalization. We illustrate this difference in Figure~\ref{fig:GlobalCoding}. In the first row, we visualize a deformation between a pair of human shapes, a category on which NJF was trained. In the second row, we evaluate the same model on a pair of cat shapes. For fairness, our LJN method \textit{was also trained on human shape pairs}, yet unlike NJF, LJN produces near-perfect deformation for shapes outside the trained category.

\section{Conclusion, Limitations and Future Work}

This paper presented LJN, a data-driven framework that efficiently estimates deformations by computing high-quality Jacobian fields from coarse inputs. Unlike the typical global encoding for representing deformations, we showed that constructing appropriate local signals is generally sufficient to learn robust deformations. These input signals constrain the neural network to learn deformation within a small neighborhood while leveraging the Poisson system for global coherence. This approach makes LJN data-friendly and robust in cross-category generalization, with both supervised and unsupervised pipelines applicable to various deformation-based tasks. While tailored for detailed deformations, LJN can still produce implausible results such as shrinkage of volume or failure to produce sharp bending. Future work could include integrating physics-based energies into the data-driven realm for more realistic deformations. Another avenue is to explore better approximations to deformation spaces, such as constructing input signals from eigenfunctions of the Discrete Shell-Operator~\cite{Tamstorf2013}.

\small
\section{Acknowledgement}
Parts of this work were supported by the ERC Starting Grant 758800 (EXPROTEA), ERC Consolidator Grant 101087347 (VEGA), ANR AI Chair AIGRETTE, as well as gifts from Ansys and Adobe Research. We also gratefully acknowledge the support of NVIDIA Corporation to the University of Milano-Bicocca, PRIN 2022 project `GEOPRIDE' and MUR under the grant ``Dipartimenti di Eccellenza 2023-2027''. This work was performed using HPC resources from GENCI–IDRIS (Grant 2023-AD011013104). We thank Guillaume Coiffier for insightful discussions. 
\bibliographystyle{ACM-Reference-Format}
\bibliography{sample-base}

\pagebreak
\clearpage
\twocolumn[
    \begin{center}
        {\Huge Appendix}
        \vspace{2cm}
    \end{center}]
\appendix


\subsection{Overview}
In this document, we provide additional details pertaining to our work. We provide an overview of all symbols used throughout the paper in Table~\ref{tab:Symbols} for expositional clarity. In Table~\ref{tab:HandleBased}, we provide the details of the dataset used in our handle-based deformation. Finally in Table~\ref{tab:UnsupSmal}, we provide quantitative results of unsupervised deformation and mapping on the SMAL dataset.

\begin{table}[b]
\centering
\resizebox{0.8\columnwidth}{!}{%
\begin{tabular}{@{}|l|ll|l|l|@{}}
\toprule
Object  & \multicolumn{2}{l|}{Symmetric Dirichlet} & \#V, F    & \#Handle \\ \midrule
        & \multicolumn{1}{l|}{ARAP}     & Ours     &           &          \\ \midrule
Pig     & \multicolumn{1}{l|}{0.26}     & 0.05     & 4K,8K     & 330      \\ \midrule
Hand    & \multicolumn{1}{l|}{5.7}      & 2.9      & 700, 1.3k & 72       \\ \midrule
Eiffel  & \multicolumn{1}{l|}{2.4}      & 0.93      & 3k, 6.1k  & 21       \\ \midrule
Chicken & \multicolumn{1}{l|}{10.5}     & 6.8      & 8k, 16k   & 400      \\ \midrule
Head    & \multicolumn{1}{l|}{1.9}      & 21.1     & 5k, 8k    & 63       \\ \midrule
Cat    & \multicolumn{1}{l|}{7.8}      & 4.8     & 7k, 14k    & 800       \\ \midrule
Buck    & \multicolumn{1}{l|}{6.2}      & 8.3     & 5k, 8k    & 2300       \\ \midrule
Horse    & \multicolumn{1}{l|}{15.6}     & 3.8   & 5k, 8k    & 600       \\ \bottomrule
\end{tabular}%
}
\caption{Quantitative comparison for handle-based deformation. We provide the quantitative symmetric Dirichlet scores denoting the smoothness in the deformation alongside details of the mesh used.}
\label{tab:HandleBased}
\end{table}

\begin{table}[b]
\centering
\centering
\centering
\resizebox{0.7\columnwidth}{!}{%
\begin{tabular}{|l|c|c|c|c|}
\hline
\textbf{Method}                             & \textbf{Geod} & \textbf{Inv} & \textbf{DirE} & \textbf{Cov} \\ \hline
W/o $\mathcal{L}_{J}$                       & 5.5           & 12.5         & 17.1          & 59.6        \\ \hline
$\alpha_6, \alpha_7 = 0$                    & 5.3           & 11.9         & 15.4          & 59.6        \\ \hline
$\alpha_7 = 0$                              & 9.7           & 11.7         & 25.2          & 53.2        \\ \hline
Ours                                        & 5.0           & 10.6         & 9.8           & 60.4        \\ \hline
\end{tabular}%
}

\caption{Quantitative result for unsupervised deformation on the SMAL dataset. We ablate the different terms used in our unsupervised loss function $\mathcal{L}_{un}$(c.f Eqn 13) over the SMAL dataset.}
\label{tab:UnsupSmal}
\end{table}

\begin{table}[b]
\centering
\centering
\centering
\resizebox{\columnwidth}{!}{%
\begin{tabular}{@{}|l|l|@{}}
\toprule
Symbols                       & Description                                                   \\ \midrule
$\mathcal{S}_1$               & Source Shape                                                  \\ \midrule
$\mathcal{S}_2$               & Target Shape                                                  \\ \midrule
$\mathcal{V}_i$               & Vertex corresponding to $\mathcal{S}_i$                       \\ \midrule
$\mathcal{F}_i$               & Face corresponding to $\mathcal{S}_i$                         \\ \midrule
$\Delta$                      & Laplace Beltrami Operator                                     \\ \midrule
$\nabla$                      & Discrete gradient operator $\mathbb{R}^{|3F|\times|V|}$       \\ \midrule
A              & Area of each face, written as diagonal matrix $\mathbb{R}^{|3F|\times|3F|}$         \\ \midrule
M              & Voronoi-area of each vertex, written as diagonal matrix $\mathbb{R}^{|V|\times|V|}$ \\ \midrule
$\Psi$                        & Eigenbasis of Laplace operator                                \\ \midrule
$\mathcal{I}$                 & $\mathbb{R}^{|V|\times|F|}$ vertex-face incidence matrix      \\ \midrule
$\mathcal{H}$                 & $\mathbb{R}^{|F|\times|F|}$ face-face incidence matrix        \\ \midrule
$\bar{\mathcal{V}}$           & Projection of $\mathcal{V}_i$ to the LBO eigenbasis           \\ \midrule
$\mathcal{E}_k$               & Non-orthonormal frame of face k                               \\ \midrule
$\mathbf{E}_k$                & Rewriting $\mathcal{E}_k$ across all faces in matrix form     \\ \midrule
$\varphi_{12}$ & Continuous map between $\mathcal{S}_1$ and $\mathcal{S}_2$                          \\ \midrule
$\Pi_{12}$                    & Vertex-Vertex map between $\mathcal{S}_1$ and $\mathcal{S}_2$ \\ \midrule
$\Pi^{*}_{12}$                & Refined vertex-vertex map                                     \\ \midrule
$\tilde{\Pi}_{12}$            & Soft pointwise map                                            \\ \midrule
$C_{21}$                      & Functional map corresponding to $\Pi_{12}$                    \\ \midrule
$\mathbf{J}_{12}$             & Jacobian between $\mathcal{S}_1$ and $\mathcal{S}_2$          \\ \midrule
$\Theta_{12}$                 & Input signal to our network                                   \\ \midrule
$e^{i}_1$ & First edge vector of $i^{th}$ face                            \\ \midrule
$\mathcal{Q}_{12}$            & Closest rotation matrix to $\mathbf{J}_{12}$                  \\ \bottomrule
\end{tabular}%
}
\caption{Collecting all symbols used in our main paper}
\label{tab:Symbols}
\end{table}

\end{document}